\documentclass[11pt,onecolumn]{article}
\usepackage{amsmath}
\usepackage{amssymb}
\usepackage{mathptmx}
\usepackage{hyperref}
\usepackage{xcolor}
\usepackage{authblk}
\usepackage{graphicx}
\usepackage[font=small]{caption}
\usepackage[margin=0.75in]{geometry}

\newcommand{\relbnorm}{\langle|\mathbf{B}\cdot\hat{\mathbf{n}}|
                       /|\mathbf{B}|\rangle}
\newcommand{\bnormsq}{|\mathbf{B}\cdot\hat{\mathbf{n}}|^2}

\begin{document}

\title{Geometric concepts for stellarator permanent magnet arrays}

\author[1,*]{K.C.~Hammond}
\author[1]{C.~Zhu}
\author[1]{T.~Brown}
\author[1]{K.~Corrigan}
\author[1]{D.A.~Gates}
\author[1]{M.~Sibilia}
\affil[1]{Princeton Plasma Physics Laboratory, Princeton, NJ, USA}
\affil[*]{\small Email: khammond@pppl.gov}

\maketitle

\begin{abstract}
The development of stellarators that use permanent magnet arrays to shape their 
confining magnetic fields has been a topic of recent interest, but the
requirements for how such magnets must be shaped, manufactured, and assembled
remain to be determined. To address these open questions, we
have performed a study of geometric concepts for magnet arrays with the aid
of the newly developed \textsc{Magpie} code. A proposed experiment similar
to the National Compact Stellarator Experiment (NCSX) is used
as a test case. Two classes of magnet geometry are explored: curved bricks that 
conform to a regular grid in cylindrical coordinates,
and hexahedra that conform to the toroidal plasma geometry. In addition, we
test constraints on the magnet polarization. While magnet configurations 
constrained to be polarized normally to a toroidal surface around the plasma
are unable to meet the required magnetic field parameters when subject to
physical limitations on the strength of present-day magnets, configurations
with unconstrained polarizations are shown to satisfy the physics requirements
for a targeted plasma.
\end{abstract}

\section{Introduction}
\label{sect:intro}

The stellarator, a nonaxisymmetric toroidal plasma confinement device, is
a potentially attractive concept for a fusion reactor.
It offers some inherent advantages over the axisymmetric tokamak, including
the ability to run at steady state with little to no plasma current
and a lower risk of disruptions. To realize these advantages, the 
three-dimensional confining magnetic field must be carefully designed to
avoid excessive neoclassical losses \cite{mynick2006a}. 

For all optimized stellarators constructed to date, the optimized magnetic
field properties have been obtained through the design of non-planar 
coils \cite{beidler1990a,sapper1990a,anderson1995a}.
One widely-used approach for designing such coils involves defining
a toroidal \textit{winding surface} and computing the surface current 
distribution required to accommodate a desired plasma contained within the
surface \cite{merkel1987a}. The surface current distribution can then be
discretized to determine the geometry of modular stellarator coils. This is the 
underlying approach of the \textsc{Nescoil} and \textsc{Regcoil} codes 
\cite{landreman2017a}.
Recently it was shown that an equivalent mathematical approach can be employed
to calculate a distribution of magnetization within a toroidal region
enclosing the plasma \cite{helander2020a}. This magnetized region, in 
combination with
a simple (e.g., planar) coil set to provide a toroidal magnetic field component,
can in principle confine a stellarator plasma just as well as the surface
current distribution. 

While permanent magnet arrays have not been employed in stellarators to date,
they have been designed and built to provide high-precision magnetic fields
with strengths of up to 1~T or greater in many other applications, including
particle accelerators \cite{aleksandrov2014a, thonet2016a, hoffstaetter2017a}, 
electric motors \cite{zhu2001a},
free-electron lasers \cite{roberson1989a, oshea2010a}, and magnetic resonance 
imaging \cite{bluemlich2008a}. The rare-Earth magnets typically used in
such arrays can withstand background fields of up to 2.8~T at room temperature
and up to 6~T at the temperature of liquid nitrogen without 
demagnetizing \cite{benabderrahmane2012a}. Recent advances in the synthesis of
magnetic material based on iron-nitrogen compounds have attained remanent
fields of greater than 2~T \cite{wang2012a,jiang2016a}, nearly twice the level 
of rare-Earth magnets.

The incorporation of permanent magnets to supplement the magnetic field produced
by coils has the potential to simplify the construction and 
maintenance of stellarator reactors. The contributions of permanent magnets
to the three-dimensional field shaping may enable the use of simpler
coils or even eliminate the need for non-planar coils, one
of the main drivers of the cost and complexity of stellarator construction
\cite{nielson2010a,rummel2012a,bosch2013a}.
Similarly to coils, permanent magnets would require cooling and neutron
radiation shielding to prevent degradation \cite{alderman2002a}. However,
the magnets may consist of bulk material and be partitioned
into arbitrarily small modules, suggesting lower costs, less-stringent
constraints for fabrication and assembly, and easier device access 
in comparison to coils. 

While permanent magnets offer potential benefits in principle, their use in
stellarators has not been extensively studied or tested.
One central challenge in realizing a permanent magnet-based stellarator will
be to develop a magnet arrangement that meets the physical requirements
for effective plasma confinement while also being feasible to construct.
In a close analogue to traditional coil-based stellarator designs, which 
must consist of discrete coils with limitations on current density and
curvature, permanent magnet-based stellarator designs must be compatible with 
limitations on magnetization levels in present-day materials. In addition, while
arrays of magnets with arbitrary distributions of strength and polarization
direction offer many degrees of freedom toward attaining an optimal confining
field, they must ultimately be feasible to mount and enable periodic 
disassembly and reassembly for reactor maintenance.

A number of recent developments have been made toward attaining optimal
permanent-magnet distributions with realistic properties. 
\textsc{Regcoil-PM}\footnote{\href{https://doi.org/10.5281/zenodo.3934702}
{\textcolor{blue}{https://doi.org/10.5281/zenodo.3934702}}} 
uses an analogous approach to that of \textsc{Regcoil}
to compute continuous distributions of magnetization within toroidal
volumes \cite{landreman2020a}. 
Zhu et al.~\cite{zhu2020a} introduced a linear
method to design arrays of magnets with polarizations constrained to be
perpendicular to a toroidal winding surface.
The \textsc{Famus} code \cite{zhu2020b}, based on the \textsc{Focus} 
code for stellarator coil design \cite{zhu2017a},
optimizes an arbitrary distribution
of magnetic dipoles representing a discrete array of magnets.

\textsc{Famus} offers complete flexibility on the quantity and spatial 
distribution of the magnets whose dipole moment is to be optimized. 
The user must 
therefore supply a geometric arrangement that adequately represents an array of 
magnets that is feasible to construct and obeys physical limitations on 
magnetization. As one approach toward meeting this need, we have developed the 
\textsc{Magpie} (\textbf{Mag}nets, \textbf{pie}cewise) code. 
\textsc{Magpie} generates arrays of magnets with simple shapes subject to
intuitive geometric constraints. The output can be used both as input to
\textsc{Famus} to optimize the distribution of magnet strength, as well as
a starting point for engineering design of the magnets once the required
strengths are determined.  
The ability of the code to rapidly ($\lessapprox$ 30 s on a standard
PC) generate feasible magnet arrangements will enable iterations with physics
codes to attain plasmas that possess both desirable physics properties 
and magnetic field requirements that can be met with permanent magnets.

In this paper, we introduce the code and apply it to design and assess magnet
arrangements for a small-scale stellarator based on the National Compact 
Stellarator Experiment (NCSX) \cite{zarnstorff2001a,nelson2003a}.
Sec.~\ref{sect:concepts} describes the two geometric concepts for magnet arrays
that are investigated in the paper: curved bricks, which conform to a grid in 
cylindrical coordinates, and quadrilaterally-faced hexahedra, which conform
to the geometry of a smooth toroidal surface around the plasma.
Sec.~\ref{sect:perp_axis} describes a study of hexahedral configurations with 
the simplifying constraint that all magnets must be locally perpendicular to
the toroidal limiting surface. The results indicate that such a constraint
cannot be achieved with the limitations on magnet strength in present-day
materials. Sec.~\ref{sect:free_axis} describes studies of hexahedron and
brick configurations in which the polarization direction of each magnet is
a free optimization parameter. With this additional degree of freedom,
it is possible to attain magnet configurations that meet the physics 
requirements for the target plasma equilibrium. We discuss the differences
between solutions employing hexahedral and brick magnets. In 
Sec.~\ref{sect:restrictions}, we demonstrate that both geometries admit viable 
solutions even after removing magnets that would collide with access ports 
foreseen for NCSX. In addition, the effects of requirements on 
spacing between magnets for mounting structures are quantified.

\section{Geometric concepts}
\label{sect:concepts}

As with stellarator coil design, the problem of specifying a magnet array to 
confine a three-dimensional plasma is fundamentally ill-posed 
\cite{landreman2017a}, and any given plasma configuration will admit many
possible solutions. In this study, we explore two classes of solutions that
represent different approaches to the magnet array design. In one
approach, the magnets all conform to a regular grid; in the other, the magnets
are are given custom shaping to conform closely to the toroidal plasma 
geometry. 

\subsection{Curved bricks}
\label{ssect:cbricks}

The grid-conforming magnets in considered here have the geometry of curved
bricks.
The shape of each brick may be fully specified by six parameters, 
consisting of lower and upper limits on the radial coordinate $r$, vertical 
coordinate $z$, and azimuthal (toroidal) angle $\phi$. The arrays of 
curved bricks studied to date have been further constrained such that each
brick in the array has identical dimensions in each coordinate. As such,
the arrays of bricks would conform to a regular rectangular grid in the space
of the cylindrical coordinates. In real space, the bricks are arc-shaped
extrusions whose cross-sections are identical rectangles and whose arc lengths
and curvature radii vary linearly with the radial coordinate.

\begin{figure*}
    \begin{center}
    \includegraphics[width=0.95\textwidth]{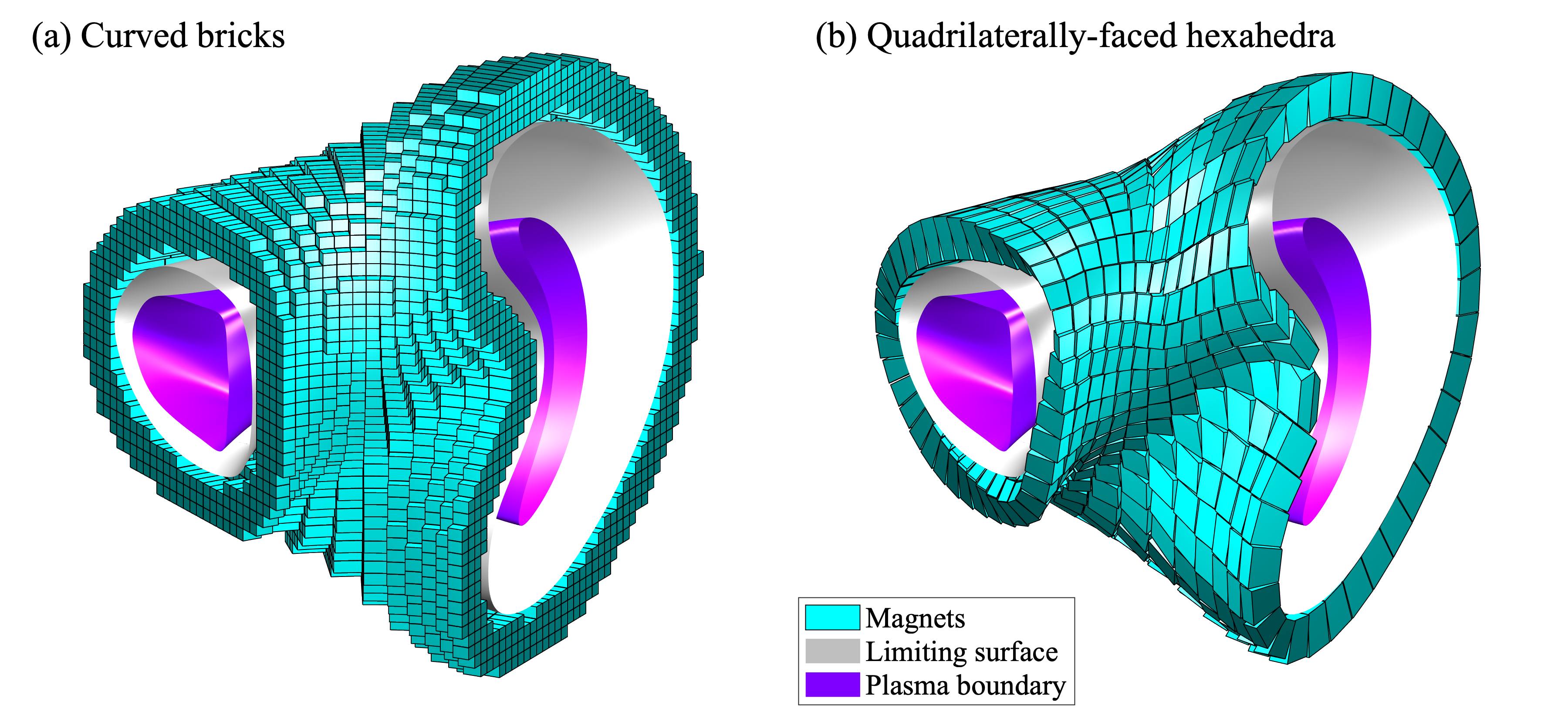}
    \caption{Renderings of arrays of (a) curved-brick magnets and 
             (b) hexahedral magnets designed for an NCSX-like plasma,
             viewed from the inboard side.
             The magnets, inner limiting surface (coincident with the vacuum 
             vessel in these cases), and the plasma boundary are shown for
             one half-period of the configuration, which has three field 
             periods in total. Both arrays have radial extents of 20 cm
             and are constructed on grids with 12 cells per half-period in
             the toroidal dimension. 
             }
    \label{fig:schematic_rendering}
    \end{center}
\end{figure*}

Fig.~\ref{fig:schematic_rendering}a shows an example configuration of curved
bricks built around an NCSX-like plasma equilibrium.
To construct a brick array for an experimental configuration, one first
defines the brick size, gap spacing, and array bounds for each cylindrical 
dimension. From this initial grid, all bricks not entirely within a specified 
permissible volume are eliminated. This volume is defined as the space between 
two toroidal \textit{limiting surfaces}: an inner surface that encloses the 
plasma, and an outer surface that encloses the inner surface. In the 
configurations we have explored to date, the inner limiting surface has 
conformed to the experimental vacuum vessel, and the outer limiting surface has 
been the locus of points whose closest distance to the inner surface in their 
respective poloidal plane is equal to a given \textit{radial extent}.

One advantage of the curved-brick concept is its geometric simplicity.
Since all bricks at a given radial position have the same shape, configurations
of curved bricks will have relatively few unique magnet geometries. 
We note, however, that 
magnets of the same shape within an array will likely need to have many
different polarization directions, dependent on their location relative to
the plasma. 

\subsection{Quadrilaterally-faced hexahedra}
\label{ssect:qhex}

\begin{figure}
    \begin{center}
    \includegraphics[width=0.5\textwidth]{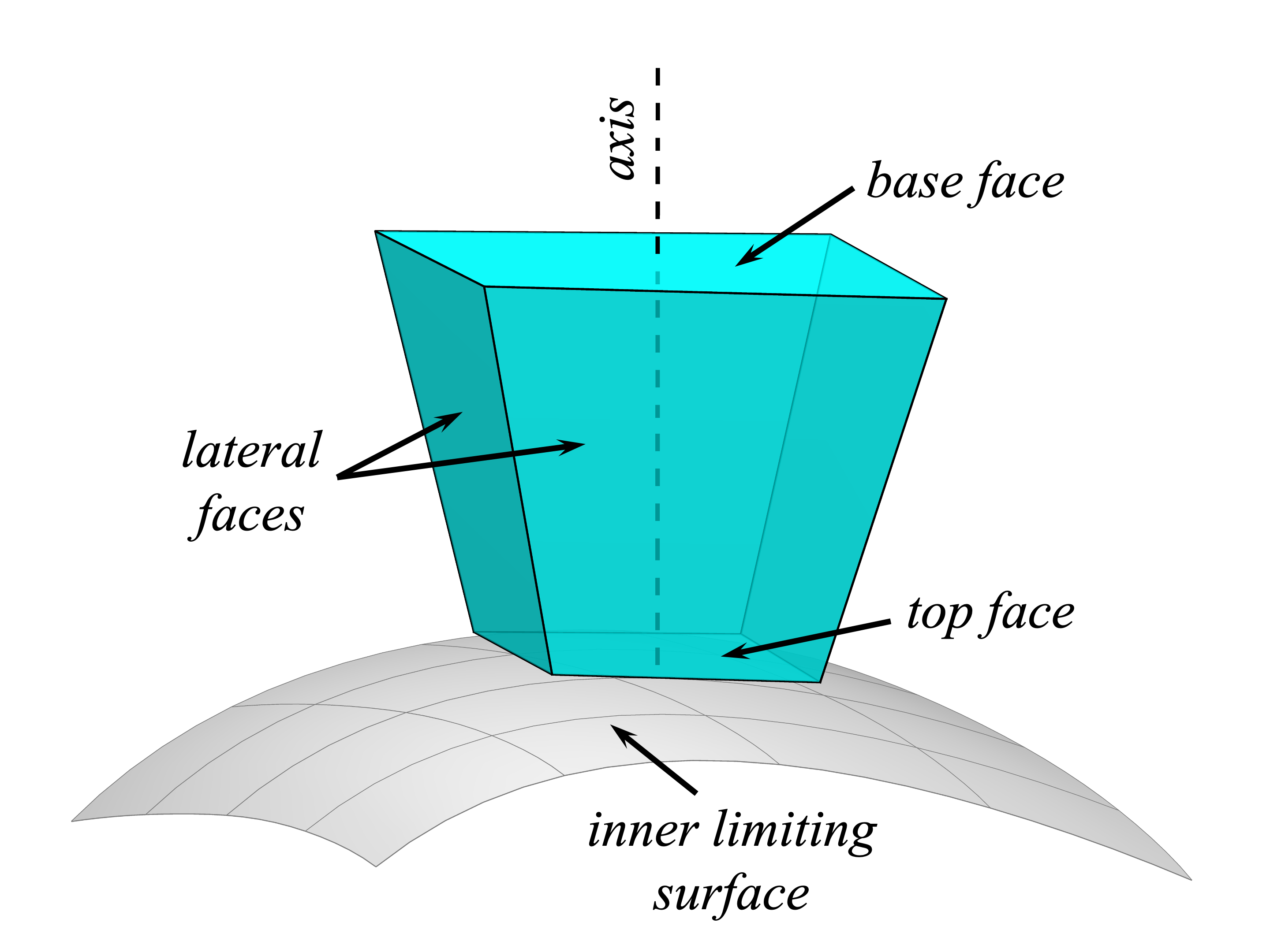}
    \caption{Schematic diagram of a quadrilaterally-faced hexahedron, labeled
             according to the terminology used in this paper. The inner
             limiting surface is also shown for reference.}
    \label{fig:qhex_diagram}
    \end{center}
\end{figure}

In the second geometric concept explored in this work, the magnets are 
constructed to conform closely to the plasma geometry. The magnets are placed
in an irregular grid that wraps around the inner limiting
surface and is partitioned into cells in the poloidal and toroidal dimensions.
Each cell contains a magnet in the form of a quadrilaterally-faced hexahedron.
A diagram of a hexahedral magnet is shown in Fig.~\ref{fig:qhex_diagram}.
Each of the four \textit{lateral} faces of 
a hexahedral magnet are parallel to the faces of the magnets in their 
respective adjacent grid cells. The \textit{top} face (closest to the inner
limiting surface) and \textit{base} face (furthest from the inner limiting 
surface) are constrained to be parallel to
one another. Each hexahedron is associated with a characteristic \textit{axis},
a line that is perpendicular to the base and top faces. The hexahedra are
positioned and aligned such that their axes are locally perpendicular to the
limiting surface.

An exemplary array of quadrilaterally-faced hexahedra is shown in 
Fig.~\ref{fig:schematic_rendering}b. As demonstrated in this layout,
the hexahedral geometry allows the magnets to fill the
volume between the two limiting surfaces. Specifically, in regions where
the inner limiting surface is convex (as is the case, for example, in
Fig.~\ref{fig:qhex_diagram}), the cross-section of the magnet tapers along
the axis from the base face to the top face.
In regions where the inner limiting surface is concave, on the other hand, 
the magnet's cross-section expands along its axis.
This geometric concept is, in effect, a three-dimensional
generalization of the assemblies of trapezoidal magnets in Halbach cylinders
employed in particle accelerators and other applications \cite{halbach1980a}.

In some grid cells in concave regions of the limiting 
surface, in particular on the inboard side near the ``bean-shaped'' 
plasma cross-section on the right of Fig.~\ref{fig:schematic_rendering}b,
additional hexahedra are added to fill gaps between the original magnet layer
and the inner limiting surface. A close-up view of a group of these additional
hexahedra is shown in Fig.~\ref{fig:upper_qhex}.

The concept of an array of magnets with axes oriented perpendicular to a
toroidal surface was motivated by the study in Ref.~\cite{zhu2020a}, which 
demonstrated the physical feasibility of plasma confinement with a magnetized 
layer whose magnetization is locally perpendicular to a toroidal surface 
surrounding the plasma. The hexahedra are a discrete implementation of such a 
layer, assuming that the magnets are each polarized along their respective
axes.

While the geometry of the hexahedral arrays is more complex than that of the
curved bricks, they could in principle offer a degree of simplicity in 
their polarization requirements. If the magnets are all polarized along
their axes (\textit{i.e.} perpendicular to the base and top faces), they may be 
cut from slabs of magnetic material with uniform
perpendicular polarization, thereby potentially reducing manufacturing 
complexity. Magnets with arbitrary polarization directions, by contrast,
would likely require more time for machining and result in more discarded 
material.

\begin{figure}
    \begin{center}
    \includegraphics[width=0.5\textwidth]{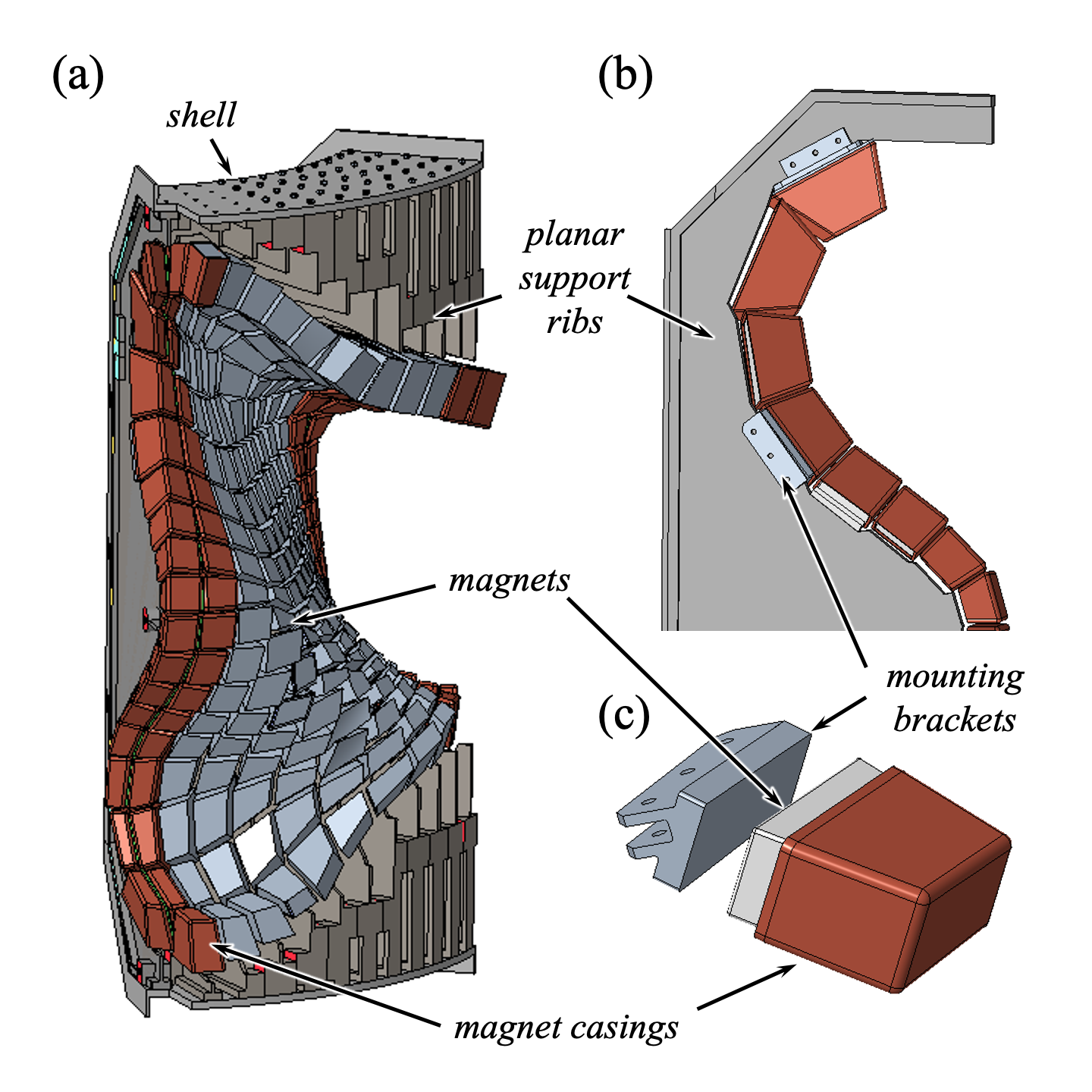}
    \caption{CAD renderings illustrating a preliminary concept for mounting
             a subset of hexahedral magnets from a variant of the 
             set depicted in Fig.~\ref{fig:schematic_rendering}b. 
             (a) Assembly of magnets fixed to planar support ribs, which are
                 in turn bolted to an external shell. Some poloidal rows of 
                 magnets are shown with stainless steel casings (brown). The 
                 sub-assembly shown here is viewed from the outboard side; 
                 the support structures are on the inboard side.
             (b) Upper portion of one of the planar support ribs with 
                 encased magnets, two of which include brackets for mounting.
             (c) Exploded view of a mounting bracket, magnet, and casing.
                 The casing is foreseen to be welded to the mounting bracket. }
    \label{fig:sas_ribs}
    \end{center}
\end{figure}

While the magnet axes are constrained to align with normal vectors to the 
limiting surface, the hexahedral concept still allows some freedom to choose
the locations of the bases of the magnets within the grid to be convenient
for mounting. In the hexahedral configurations explored to date, the 
magnet bases have been specified such that the bases in each poloidal row of
the grid align to a constant toroidal angle. Thus, while the axes in a
given grid row emerge from the bases in different directions dependent on 
the geometry of the inner limiting surface, the bases may all be mounted on 
planar external structures as shown in Fig.~\ref{fig:sas_ribs}.

Further details on the algorithm for designing the hexahedral arrays are 
given in Appendix~\ref{app:qhex_construction}.

\section{Magnet arrays with normal polarization}
\label{sect:perp_axis}

The two magnet concepts introduced in Sec.~\ref{sect:concepts} will be 
evaluated for suitability for a proposed variant of the NCSX stellarator.
In each case, the magnets will replace the modular, non-planar coils of NCSX.
The toroidal field is supplied by planar toroidal field (TF) coils. 
The target equilibrium is based on an optimized plasma configuration for NCSX.
In absence of the modular coils, which would have contributed a substantial
portion of the toroidal field, the capabilities of the TF coils alone limit
the magnetic field strength 0.5~T. This is about one-third of 
the originally envisioned field \cite{zarnstorff2001a}.
The major characteristics of the target 
plasma equilibrium are summarized in Table~\ref{tab:ncsx}.

\begin{table}
    \begin{center}
    \begin{tabular}{|c|c|}
    \hline
    Major radius                      & 1.44 m  \\
    Minor radius                      & 0.32 m  \\
    $|\mathbf{B}|$ (volume average)   & 0.50 T  \\
    Rotational transform on axis      & 0.35    \\
    $\beta$ (volume average)          & 4.1\%   \\
    Toroidal current                  & 57 kA   \\
    \hline
    \end{tabular}
    \caption{Key characteristics of the NCSX-like target plasma equilibrium
             used for the magnet studies.}
    \label{tab:ncsx}
    \end{center}
\end{table}

\subsection{Optimization procedure}
\label{ssect:optimization}

To determine the suitability of a given magnet array for this target plasma,
we employed the \textsc{Famus} code, which optimizes the dipole moment of each
magnet in order to match the required magnetic field for the target 
plasma \cite{zhu2020b}. Geometric magnet arrays were considered usable if
a solution with sufficiently low residual field error could be found, subject
to constraints on magnetization. For these optimizations, each
magnet was represented as an idealized point dipole located at the magnet's
centroid. The maximum allowable dipole moment was capped
at the magnet's volume times a designated maximum bulk magnetization 
$M_\text{max}$. For these studies, we have used 
$M_\text{max}=1.1~\text{MA}/\text{m}$, corresponding to the level presently
attainable in rare-Earth magnets \cite{benabderrahmane2012a}.
We note that, while the optimization procedure employed here produces
arrays with continuous ranges of magnetization strength, 
an optimized magnet with a magnetization less than $M_\text{max}$ would
be implemented in practice as a magnet with a magnetization of $M_\text{max}$
but with its dimensions reduced in order to preserve the optimized dipole 
moment.

Optimizations proceeded in two steps, both of which involved the minimization
of objective functions with a quasi-Newton method as described in 
Ref.~\cite{zhu2020b}. In the 
first step, the magnets, represented as dipoles, were initialized with a 
dipole moment of zero. The dipole moments were then optimized with the sole
objective of minimizing the integral of the squared normal component of the 
magnetic field, $\bnormsq$, over the boundary of the
target plasma. Here, $\mathbf{B}$ is the net magnetic field, including 
contibutions from the magnets, toroidal field coils, and plasma, and 
$\hat{\mathbf{n}}$ is the unit normal vector on the plasma boundary.
In the second step, the dipole moments
output from the first step were refined in an optimization with two 
minimization objectives: the integral of $\bnormsq$ and the sum of the squares 
of the dipole 
moments of each of the magnets. The goal of this refinement
was to concentrate the magnetization within a smaller number of magnets within
the arrangement and ultimately reduce the required volume of
magnetic material for the array. 

\subsection{Scan of magnet layer thickness}
\label{ssect:qhex_r_scan_normal}

The first magnet arrays tested for this target plasma consisted of 
hexahedral magnets (Sec.~\ref{ssect:qhex}) with polarizations normal to the 
inner limiting surface. As such, the directions of the dipole moments in the
\textsc{Famus} optimizations were fixed to be along the axes of the hexahedra, 
whereas the magnitudes were free to vary within the bounds permitted 
by $M_\text{max}$. The polarity of the dipoles, \textit{i.e.} whether their
moments were parallel or antiparallel to the hexahedral axes, were also free to 
vary. The hexahedra were constructed around an inner limiting surface with the 
dimensions of the NCSX vacuum vessel \cite{nelson2003a,dahlgren2005a}, which 
we note is not the same as the winding surface used to constrain the NCSX 
coils \cite{pomphrey2001a}. 
A minimum gap spacing of 2 cm was enforced between 
the magnets and the limiting surface.
The minimum spacing between parallel faces of adjacent hexahedra was 0.5~cm,
representing a highly optimistic estimate of the amount of room required between
magnets for support infrastructure.
In addition, each hexahedron was subdivided into 10 radially-arranged slices
to enable greater spatial resolution of the required dipole moment distribution
within the volume of the array.
This is instructive, as each hexahedron in the array will likely consist of a 
subassembly of many smaller magnets.

\begin{figure*}
    \begin{center}
    \includegraphics[width=\textwidth]{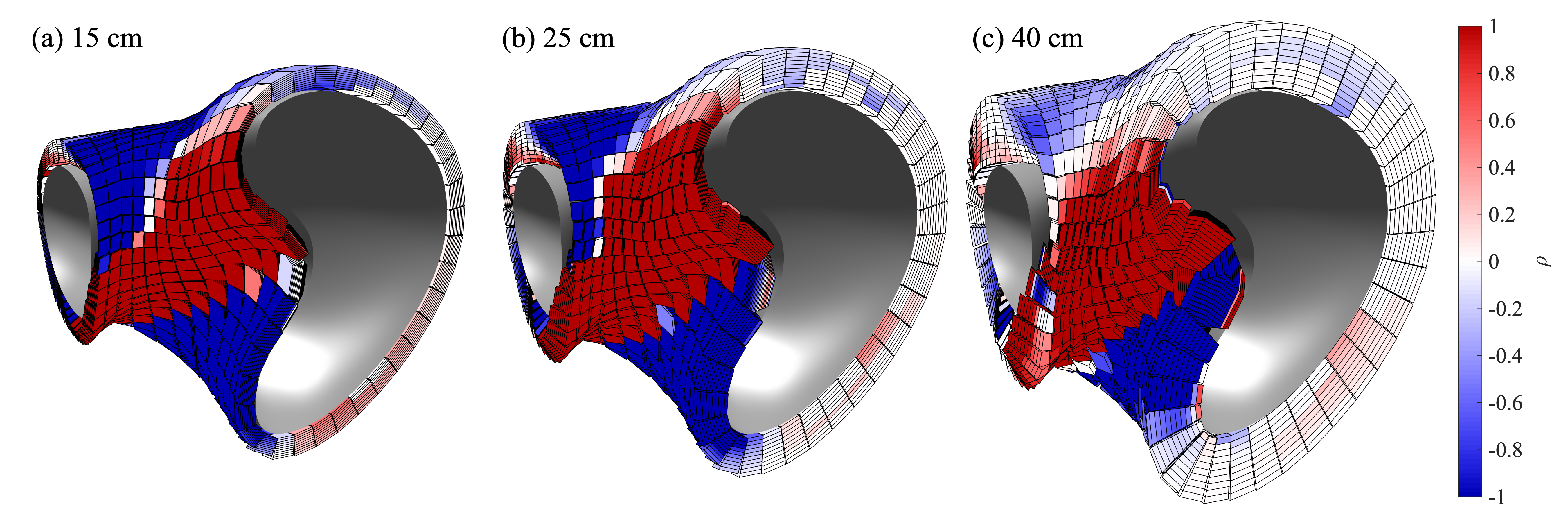}
    \caption{Renderings of arrangements of quadrilaterally-faced hexahedra
             with radial extents of (a) 15 cm, (b) 25 cm, and (c) 40 cm.
             The color scale corresponds to the value of $\rho$ obtained
             from an optimization in which the dipole moment's direction
             was fixed to be along the axis of the hexahedra, locally normal
             to the inner limiting surface. Positive values of $\rho$ correspond
             to the outward direction.}
    \label{fig:magnets_qhex_norm_scan}
    \end{center}
\end{figure*}

Each of the arrays had a distinct, uniform value of the radial extent parameter,
ranging from 15~cm to 40~cm. Three of these arrays, having radial extents
of 15~cm, 25~cm, and 40~cm, respectively, are shown in 
Fig.~\ref{fig:magnets_qhex_norm_scan}.
The magnets are colored according to the optimized value of the 
\textit{density} $\rho$, defined as $|\mathbf{m}|/M_\text{max}V$, where 
$\mathbf{m}$ is the optimized dipole moment and $V$ is the magnet volume.
In this figure, $\rho$ is given a positive sign if the 
optimized dipole moment points away from the limiting surface and 
a negative sign if the moment points toward the limiting surface.
As indicated in the renderings, magnets on the inboard side tend to require
greater strengths regardless of the array thickness. This is consistent with
the findings in Ref.~\cite{zhu2020b} and indicates that many of the magnets
on the outboard side could be removed in further refinements of the arrays
without much loss to the attainable magnetic field accuracy.

The general objective of the scans of magnet thickness was to identify
the minimum thickness (correlated with magnet quantity) needed to produce
the magnetic field required for the target plasma. The attainable field 
accuracy for a given magnet set is quantified here by surface average of the
fractional normal component of the net magnetic field, $\relbnorm$.
We find that values of $\relbnorm < 0.002$
tend to produce good agreement with the boundary, 
rotational transform profile, and effective ripple of the target plasma
configuration, as will be shown later. We will therefore adopt this as
an empirical criterion for determining whether a magnet arrangement can
confine the target plasma.

\begin{figure}
    \begin{center}
    \includegraphics[width=0.5\textwidth]{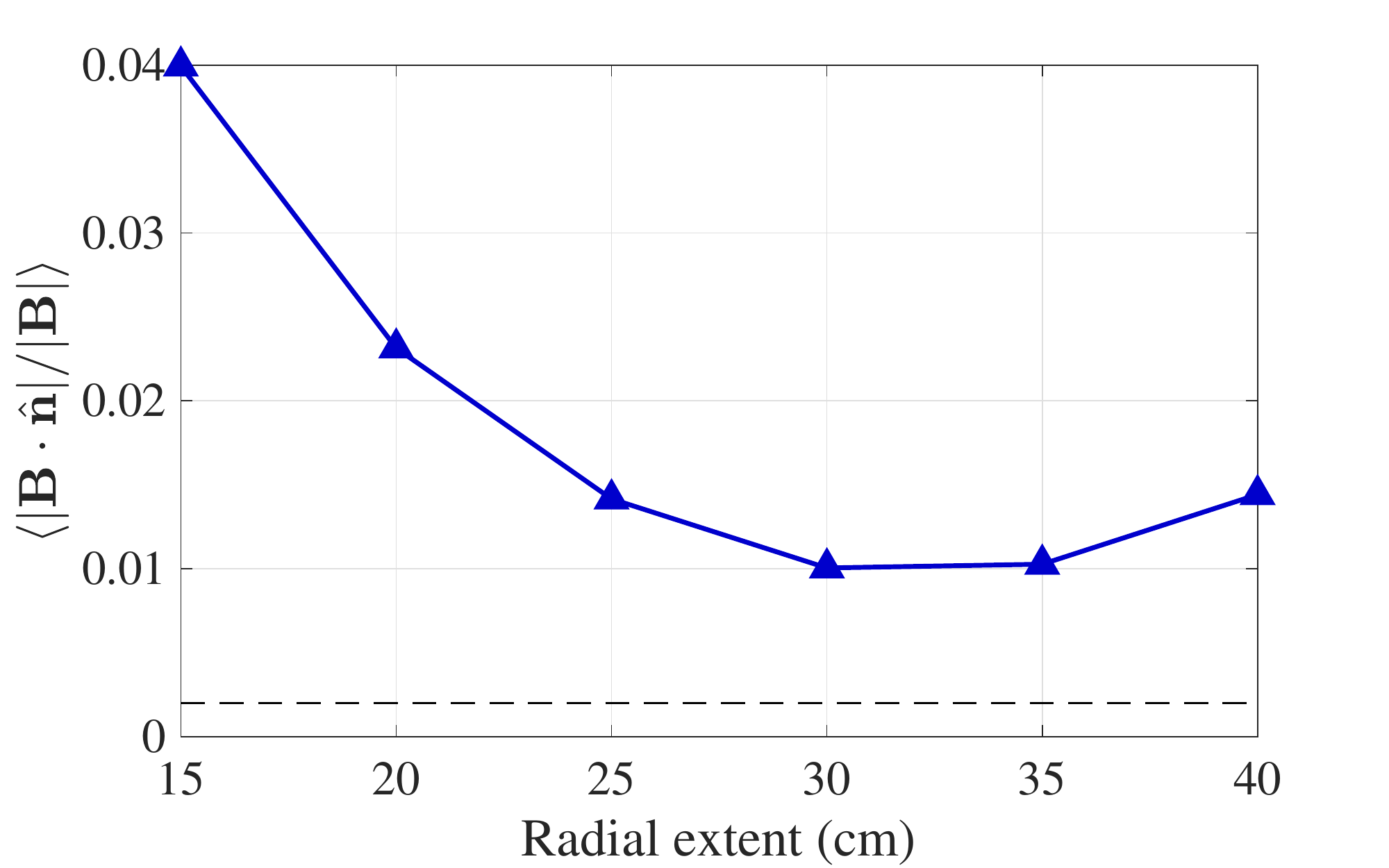}
    \caption{Values of $\relbnorm$ obtained in optimizations of hexahedral
             magnet arrangements with normal polarization. The dashed line
             indicates the criterion value $\relbnorm=0.002$.}
    \label{fig:bnorm_r_qhex_norm}
    \end{center}
\end{figure}

The values of $\relbnorm$ attained
for each of the magnet arrays in the thickness scan are shown in 
Fig.~\ref{fig:bnorm_r_qhex_norm}. As shown in the plot, none of the arrays
tested exhibited $\relbnorm<0.01$, indicating that
none were able to produce a sufficiently accurate magnetic field. The highest
field error was observed in the thinnest array (radial extent of 15~cm),
and the field error decreased substantially for the first few increments 
of the radial extent. Such a trend indicates that the thinnest arrays did not
contain enough magnetic material to counterbalance the normal components
of the field on the plasma boundary arising from the toroidal field coils and 
the plasma. However, simply adding magnets
was not sufficient to reduce the field error to acceptable levels. 
This can be understood from the fact that incrementing the radial extent
entails adding magnets that are further and further from the plasma, which 
therefore have less and less of an effect on the field at the plasma boundary.
The increase in field error in the final increment is likely due to the 
presence of larger gaps between the magnets, which tend to be necessary
to prevent overlaps between the corners of nearby magnets in thicker 
hexahedral arrays.

\subsection{Scan of maximum magnetization}
\label{ssect:qhex_M_scan_normal}

\begin{figure}
    \begin{center}
    \includegraphics[width=0.5\textwidth]{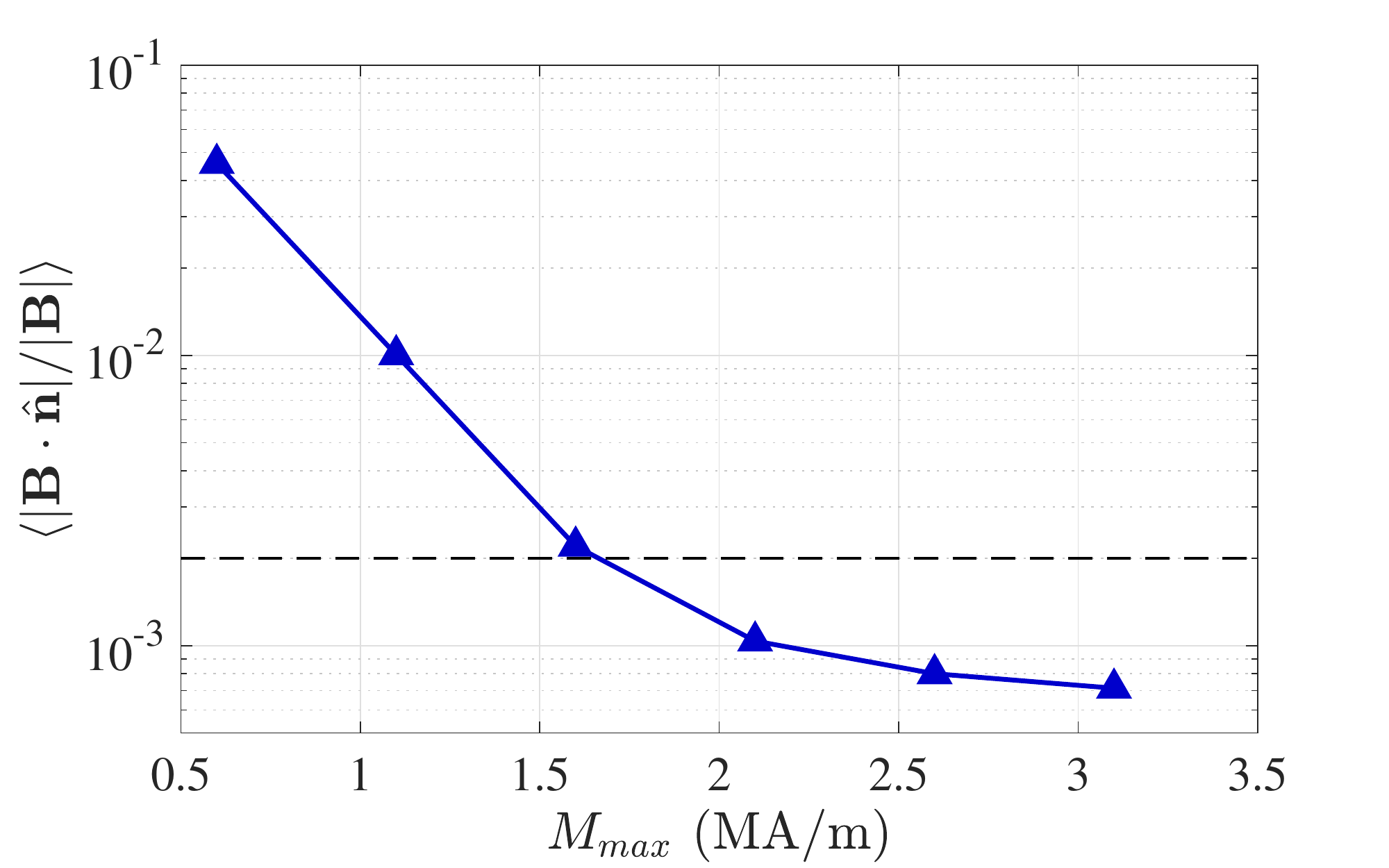}
    \caption{Values of $\relbnorm$ obtained in optimizations of hexahedral
             magnet arrangements with a radial extent of 30 cm with 
             different values of $M_\text{max}$.}
    \label{fig:bnorm_M_qhex_norm}
    \end{center}
\end{figure}

A second parameter scan with normally-polarized hexahedral magnets further
supports the above interpretation. In this scan, the radial extent was fixed
at 0.3 m, the value at which the smallest value of 
$\relbnorm$ was attained in the first scan.
The varied parameter in this case was $M_\text{max}$, which ranged from
0.6~$\text{MA}/\text{m}$ to 3.1~$\text{MA}/\text{m}$.
As indicated in Fig.~\ref{fig:bnorm_M_qhex_norm}, the field error diminished
substantially as $M_\text{max}$ increased, far below the level achievable
through increasing the magnet layer thickness with $M_\text{max}$ fixed
at 1.1~$\text{MA}/\text{m}$ (Fig.~\ref{fig:bnorm_r_qhex_norm}).

The ability of the different magnet arrays in this scan to meet the 
physics requirements for plasma confinement was investigated through
magnetohydronamic (MHD) equilibrium and neoclassical confinement calculations.
We used the VMEC code \cite{hirshman1983a,hirshman1986a} to calculate 
free-boundary MHD plasma equilibria using the external fields provided by the 
NCSX toroidal
field coils and each of the optimized magnet arrangements. In addition,
the NEO code \cite{nemov1999a} was used to calculate $\varepsilon_\text{eff}$,
a metric of neoclassical confinement. 

\begin{figure*}
    \begin{center}
    \includegraphics[width=\textwidth]{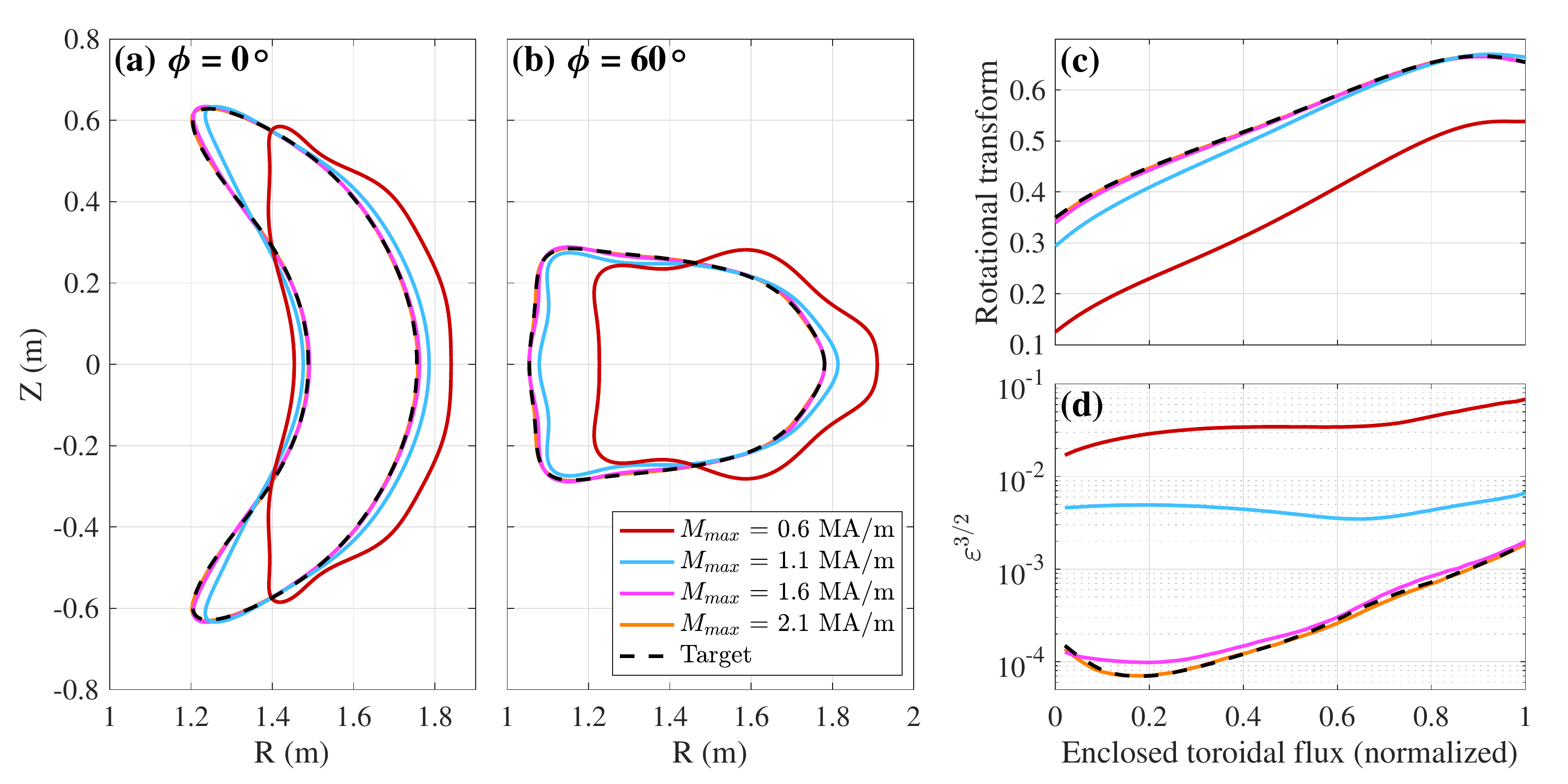}
    \caption{Results of MHD equilibrium and neoclassical transport calculations
             for plasmas confined by an arrangement of hexahedral magnets with
             differing values of $M_\text{max}$.
             (a) Plasma boundary at toroidal angle $\phi=0^\circ$;
             (b) Plasma boundary at toroidal angle $\phi=60^\circ$;
             (c) Profile of rotational transform;
             (d) Profile of $\varepsilon_\text{eff}^{3/2}$.}
    \label{fig:equil_qhex_Mscan}
    \end{center}
\end{figure*}

Results from these calculations for selected arrangements are shown in
Fig.~\ref{fig:equil_qhex_Mscan}, including the equilibrium plasma boundary,
rotational transform profile, and $\varepsilon_\text{eff}^{3/2}$ as compared
to the corresponding data for the target plasma configuration. The agreement
between the attainable and targeted plasma boundaries and parameter profiles 
for the arrangements with $M_\text{max}=0.6~\text{MA}/\text{m}$ and 
$1.1~\text{MA}/\text{m}$ was poor, particularly for the neoclassical
transport. However, all properties converged toward the target as 
$M_\text{max}$ increased, with good agreement at 
$M_\text{max}=1.6~\text{MA}/\text{m}$ and excellent agreement at
$M_\text{max}=2.1~\text{MA}/\text{m}$. Since those latter two arrangements
had values of $\relbnorm$ of $2.20\times{10}^{-3}$ and $1.04\times{10}^{-3}$
respectively, we have chosen $\relbnorm<0.002$ as an empirical criterion
for adequate magnetic field accuracy when evaluating subsequent magnet
arrangements.

While values of $M_\text{max}$ of 2.1~$\text{MA}/\text{m}$ are far above the
limitations of present-day materials, these results are instructive for the
purposes of magnet array design. While arrays of magnets with normal 
polarization can in principle confine stellarator plasmas, they require high
magnetic concentrations close to the plasma. For the target plasma explored
here, normally-polarized magnet arrays cannot generate adequate confining
fields---at least if they must be placed outside the NCSX vacuum vessel.

\section{Magnet arrays with arbitrary polarizaion}
\label{sect:free_axis}

\subsection{Hexahedral thickness scan, revisited}
\label{ssect:qhex_r_scan_free}

\begin{figure}
    \begin{center}
    \includegraphics[width=0.5\textwidth]{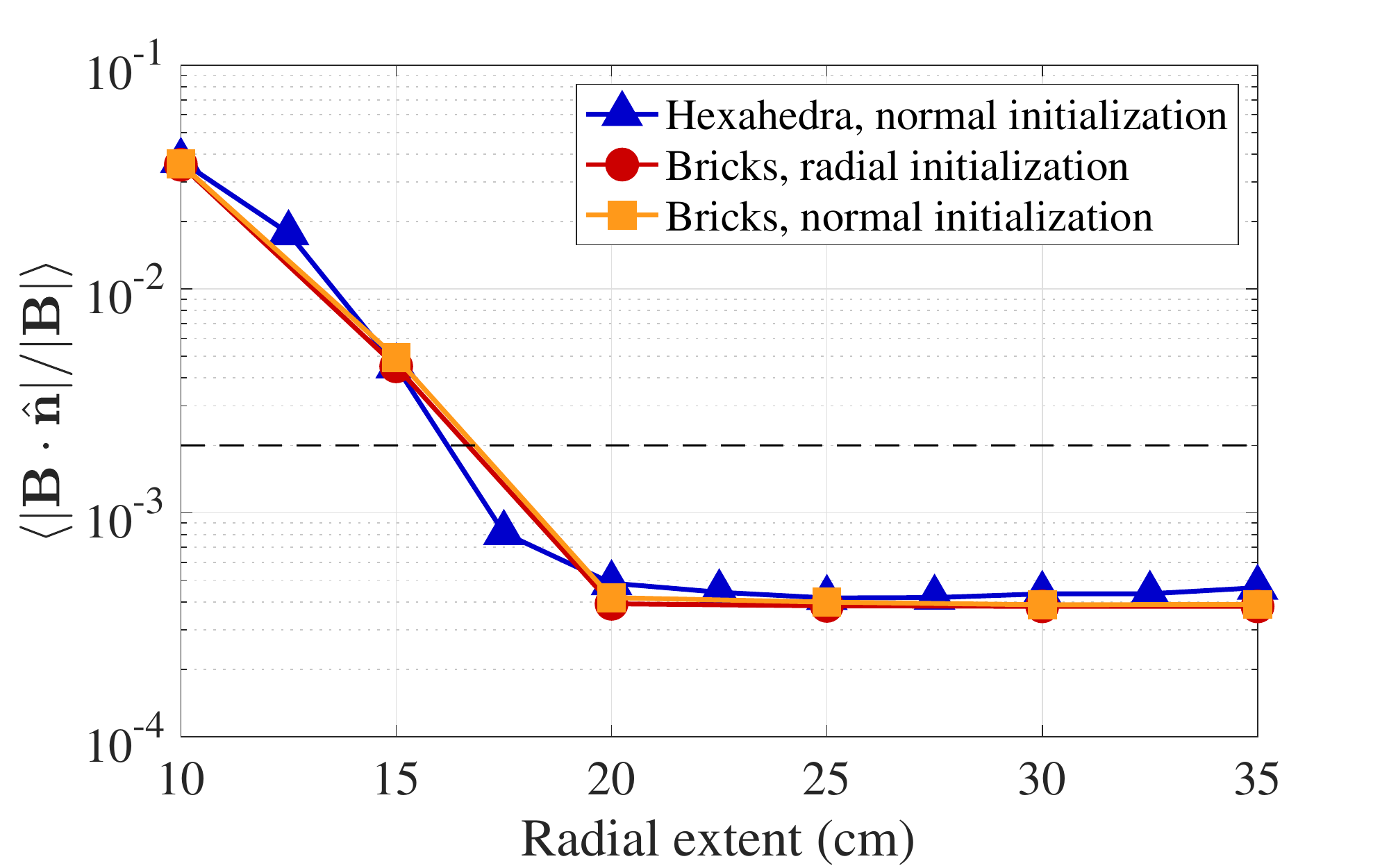}
    \caption{Values of $\relbnorm$ for different magnet arrangements for
             optimizations in which both the magnitude and direction of the
             dipole moment of each magnet were allowed to vary. The horizontal
             dashed line indicates the empirical criterion for field accuracy.}
    \label{fig:bnorm_r_free}
    \end{center}
\end{figure}

\begin{figure*}
    \begin{center}
    \includegraphics[width=\textwidth]{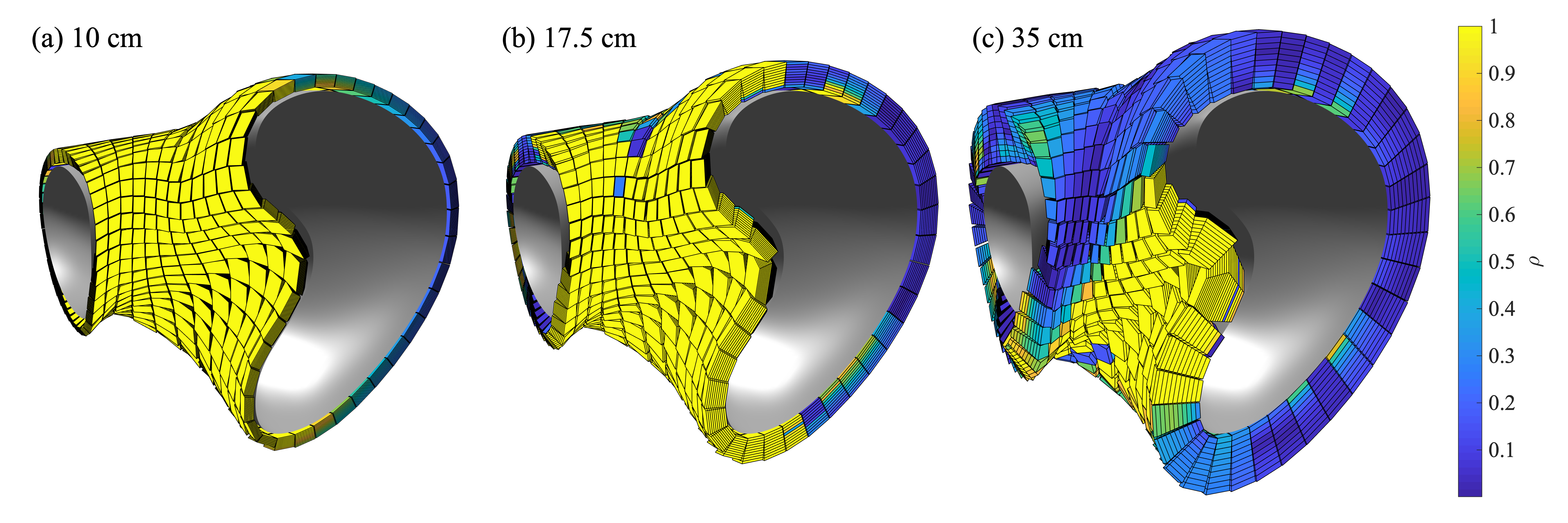}
    \caption{Renderings of arrangements quadrilaterally-faced hexahedra
             with radial extents of (a) 10 cm, (b) 17.5 cm, and (c) 35 cm.
             The color scale corresponds to the value of $\rho$ obtained
             from an optimization in which the dipole moment's direction
             could vary during the optimization.}
    \label{fig:magnets_qhex_free_scan}
    \end{center}
\end{figure*}

The set of hexahedral arrays with varying radial extents first discussed in
Sec.~\ref{ssect:qhex_r_scan_normal} was subjected to a second set of 
optimizations, this time allowing for the polarization direction to vary
relative to the hexahedrons' axes. The initial guess for the dipole moment
in each case was a vector with magnitude zero parallel to the axis of the 
corresponding hexahedron. As before, $M_\text{max}$ was set at the
rare-Earth magnet value of $1.1~\text{MA}/\text{m}$ in each case.
As shown in Fig.~\ref{fig:bnorm_r_free}, the attainable values of
$\relbnorm$ were much lower than in the fixed-axis case. In particular, 
arrangements with radial extents as low as 17.5~cm were usable in the sense
of attaining $\relbnorm<0.002$. Renderings of three arrangements from the
scan, indicating the spatial distribution of optimized $\rho$ values,
are shown in Fig.~\ref{fig:magnets_qhex_free_scan}.

\begin{figure}
    \begin{center}
    \includegraphics[width=0.5\textwidth]{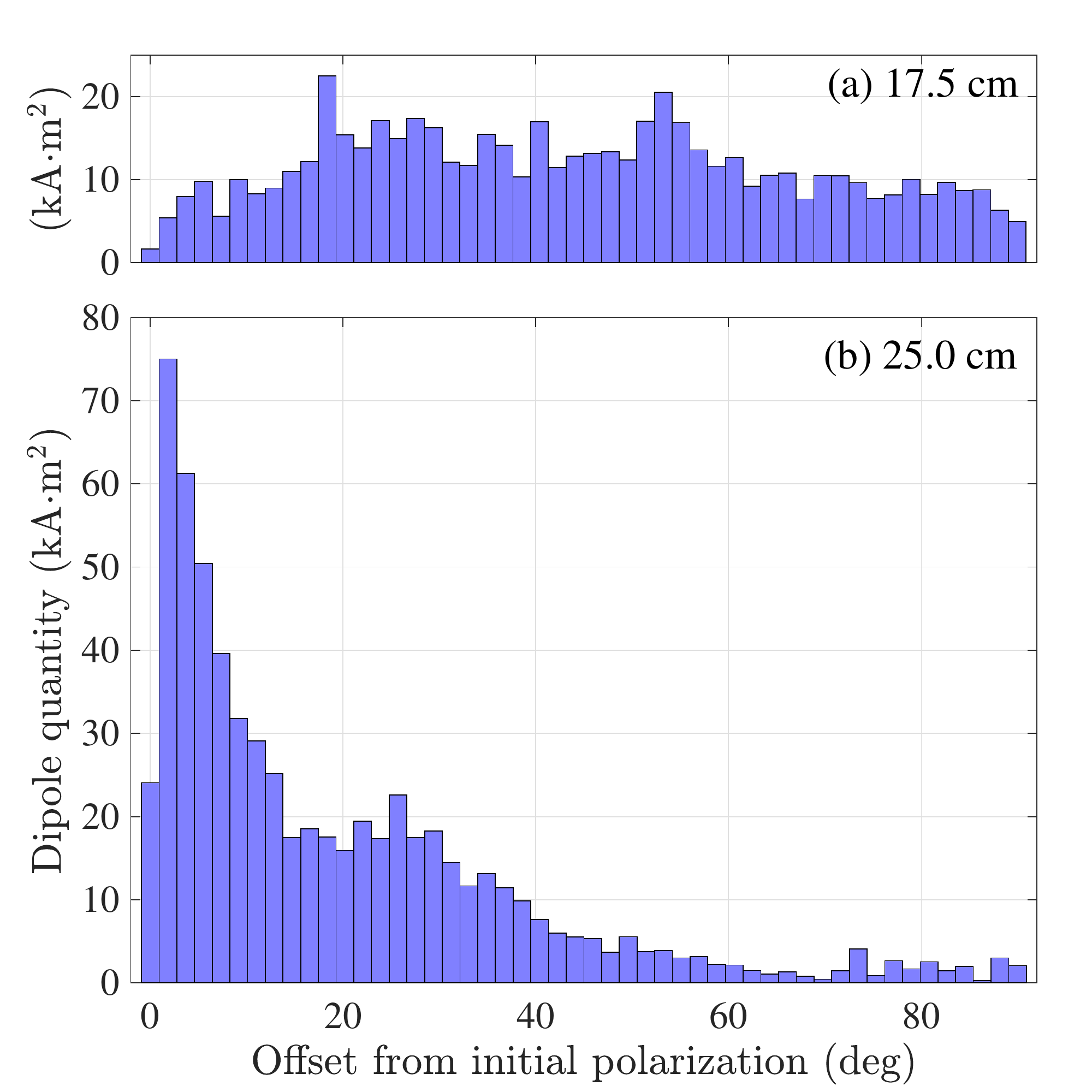}
    \caption{Distribution of the angular offset between the optimized and
             initial dipole moments for hexahedral magnets in arrangements
             with radial extents of (a) 17.5~cm and (b) 25~cm. Bin quantities
             are weighted according to the magnitudes of the optimized dipole
             moments.}
    \label{fig:qhex_angle_hist}
    \end{center}
\end{figure}

The importance of the freedom to decouple the magnetic moments from the normal
direction is illustrated in the distribution of the optimized polarization
directions. Fig.~\ref{fig:qhex_angle_hist}a shows a histogram of the 
cumulative dipole moment strength of the magnets, binned according to the
angle between the optimized dipole axis and the axis of the corresponding
hexahedral magnet body. An angle of $0^\circ$ corresponds to polarization
along the hexahedron's axis (\textit{i.e.} locally normal to the limiting 
surface) in either the positive or negative direction, whereas an angle of
$90^\circ$ corresponds to polarization perpendicular to the hexahedron's axis
(\textit{i.e.} locally tangential to the limiting surface).
The histogram corresponds to the optimized dipole moments of the arrangement
with the lowest usable radial extent of 17.5~cm. For this arrangement
the angular offset is distributed widely between normal and tangential.

Thicker magnet layers admitted distributions that were more skewed toward
the normal ($0^\circ$) orientation, as shown in Fig.~\ref{fig:qhex_angle_hist}b
for the arrangement with a radial extent of 25~cm. However, note that this
arrangement has a much greater total dipole quantity (exhibited by the sum
over all bins), indicating that the array makes less efficient use of
its magnet material in generating the required field. Furthermore, the
distribution still has a finite spread, indicating the continued necessity
for at least some magnets to have tangential components to their polarization
despite the increased thickness.

\subsection{Arrangements with curved-brick magnets}
\label{ssect:cbrick_r_scan_free}

The results in Sec.~\ref{ssect:qhex_r_scan_normal} indicate that hexahedral 
magnet arrays with the simplifying constraint of normal
polarization cannot adequately confine target plasma configuration.
Rather, the magnet arrangement must have a tangential polarization in some
regions and normal polarization in others.
Since the potential simplicity of normally-polarized hexahedral magnets
cannot be realized, the curved brick concept introduced in 
Sec.~\ref{ssect:cbricks} could be preferable to the hexahedra due to its 
relative geometric simplicity.

Accordingly, a set of curved brick arrangements with differing values of
radial extent was performed to identify possible arrangements. For each
arrangement, the poloidal cross-section of each brick was a square with side 
lengths of 4.9~cm in the radial and vertical dimensions. Each brick subtended
a toroidal angle of $4.86^\circ$. Gap spacings of 0.1~cm were maintained 
between adjacent bricks in the radial and vertical dimensions, and
$0.14^\circ$ in the toroidal dimension. 

The geometric orientation of each curved brick is independent of the
plasma geometry. As a result, the choice of initial guess for the magnetic
polarization direction is not as intuitive as for the hexahedra, which have
characteristic axes perpendicular to a toroidal surface around the plasma.
In this study, two separate optimizations were performed for each arrangement
of curved bricks. For the first optimization, the polarization direction of
each brick was initialized to be along the radial unit vector---that is,
$\cos\phi\hat{\mathbf{x}} + \sin\phi\hat{\mathbf{y}}$, where $\hat{\mathbf{x}}$
and $\hat{\mathbf{y}}$ are unit vectors in the Cartesian $x$ and $y$ directions
and the toroidal angle $\phi$ is evaluated at the brick's centroid.
For the second, the direction was initialized along a line originating from
the brick and intersecting the inner limiting surface at
a perpendicular angle. In this latter scheme, the spatial distribution 
of initial polarizations throughout the magnet volume is similar to that
of an array of hexahedra initialized along their axes.
In all cases, the dipole moment magnitude was initialized at zero and
$M_\text{max}$ was constrained not to exceed $1.1~\text{MA}/\text{m}$.

\begin{figure*}
    \begin{center}
    \includegraphics[width=\textwidth]{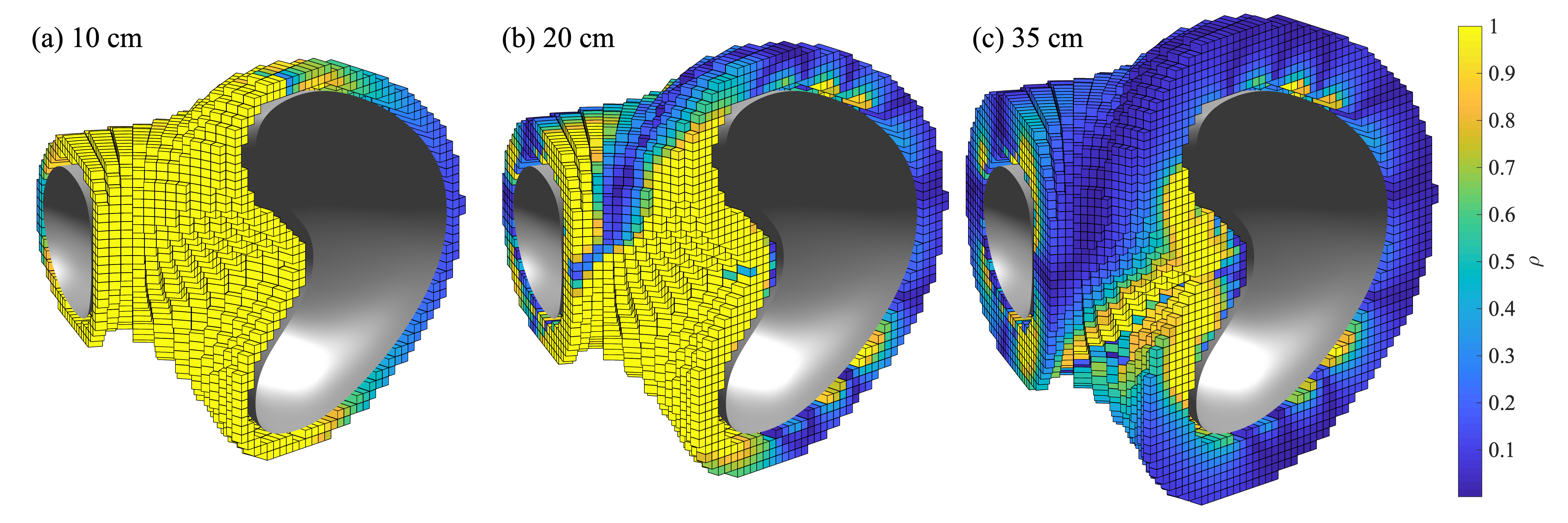}
    \caption{Renderings of arrangements curved bricks with radial 
             extents of (a) 10 cm, (b) 20 cm, and (c) 35 cm.
             The color scale corresponds to the value of $\rho$ obtained
             from an optimization in which the dipole moment's direction
             was initialized to point normally with respect to the 
             inner limiting surface.}
    \label{fig:magnets_cbrick_scan}
    \end{center}
\end{figure*}

The values of attainable $\relbnorm$ for each arrangement are plotted
in Fig.~\ref{fig:bnorm_r_free}. Note that the minimum radial extent required
to meet the criterion $\relbnorm<0.002$, approximately 20 cm, was essentially 
the same for both initialization cases, and also quite similar for the scan of 
hexahedral arrangements. Three of the arrangements, along with their optimized
$\rho$ values, are shown in Fig.~\ref{fig:magnets_cbrick_scan}. As with the
hexahedral cases, the optimized dipole moment magnitude in thicker distributions
becomes increasingly concentrated near the inboard side of the bean-shaped
plasma cross-section.

One point of contrast between the two initialization schemes can be seen in 
the effective magnet volume, $V_\text{eff}$.
We define $V_\text{eff}$ as $\sum_i^{N}|\mathbf{m}_i|/M_\text{max}$,
where $N$ is the total number of magnets in the arrangement and 
$\mathbf{m}_i$ is the optimized dipole moment of each magnet. $V_\text{eff}$
provides an estimate of the volume of magnet material required to build
the arrangement if magnets of low to intermediate optimized 
densities are substituted for smaller magnets with $\rho=1$ (\textit{i.e.}
magnetization equal to $M_\text{max}$). 
In general, lower values of $V_\text{eff}$ are desirable and, as mentioned in 
Sec.~\ref{ssect:optimization}, the second stage of each optimization sought
to reduce this quantity (more precisely, $\sum_i^{N}|\mathbf{m}_i|^2$) while
still maintaining good magnetic field accuracy. 

\begin{figure}
    \begin{center}
    \includegraphics[width=0.5\textwidth]{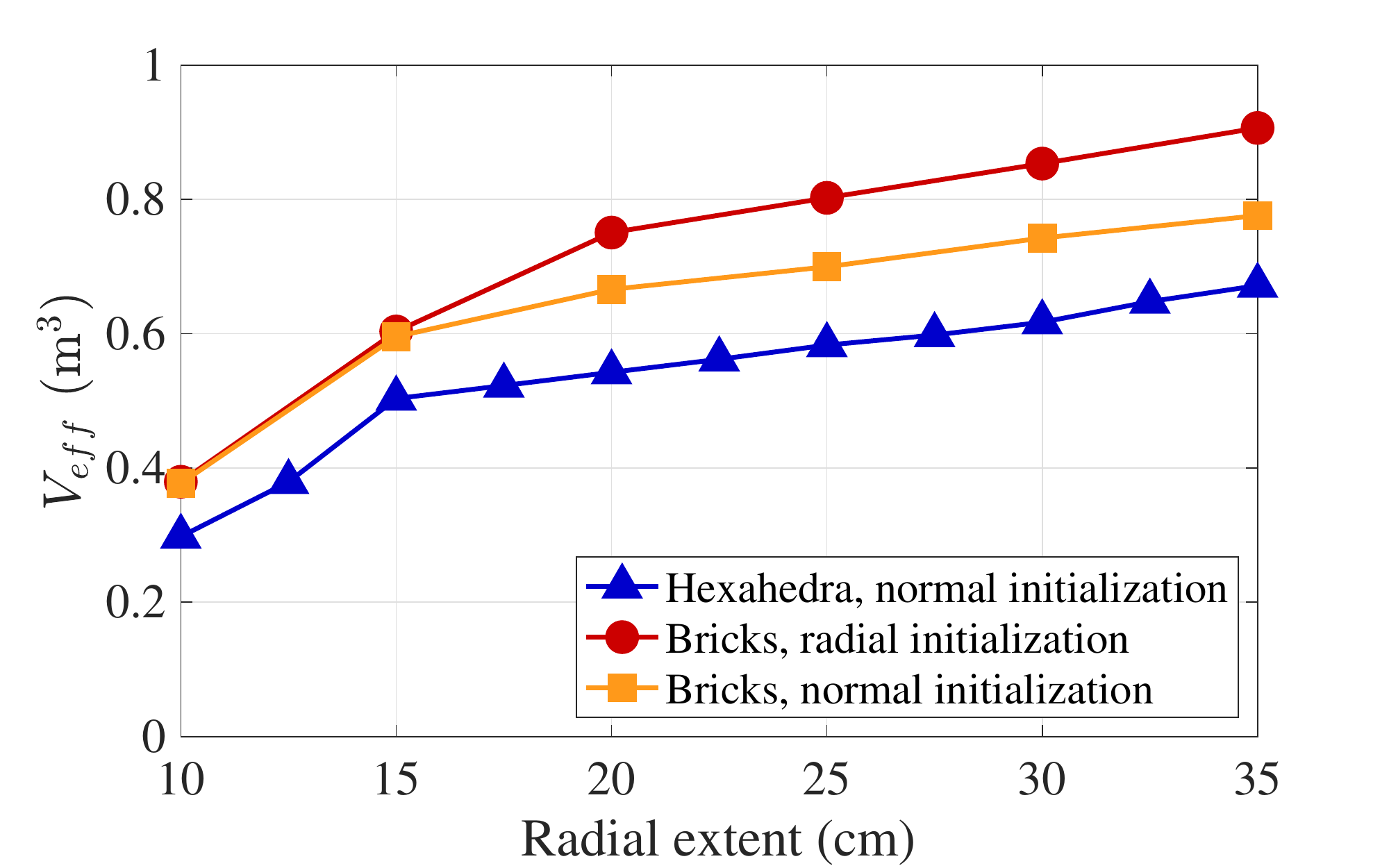}
    \caption{Effective volume per half-module for different magnet arrangements 
             for optimizations in which both the magnitude and direction of the
             dipole moment of each magnet were allowed to vary. }
    \label{fig:Veff}
    \end{center}
\end{figure}

While the attainable $\relbnorm$ appears to be fairly robust against differences
in initalization and in the geometry of the individual magnets, this cannot
be said for $V_\text{eff}$. As shown in Fig.~\ref{fig:Veff}, $V_\text{eff}$
is at least 10\% lower for magnets with the normal initialization than 
for magnets with the radial initialization, at least in the subset with 
sufficient field accuracy (radial extents of 20 cm or greater). Evidently,
the solution obtained through the optimization is sensitive to some initial
parameters, perhaps a result of local minima in the optimization 
space \cite{zhu2020b}. This sensitivity can have consequences for construction 
costs---in this case, the required magnet quantity.

Also of note are the values of $V_\text{eff}$ obtained from the free-axis
thickness scan of hexahedral magnet arrangements, also shown in 
Fig.~\ref{fig:Veff}. The hexahedral arrangements were able to
obtain similar field accuracy with substantially lower volume than 
either set of optimizations for the curved bricks. It is hypothesized that
this can be explained at least in part by the toroidally-conforming geometry
of the hexahedra. Whereas the brick arrangements exhibit many wedge-shaped
gaps near the inner limiting surface due to their rectangular cross-sections, 
the better-aligned hexahedra fill a larger fraction of the volume near the
inner limiting surface, where magnetic material would have the most influence 
on the field at the plasma boundary.

\section{Effects of spatial restrictions}
\label{sect:restrictions}

\subsection{Diagnostic ports}
\label{ssect:ports}

The magnet arrangements studied so far have all fully covered the limiting
surface, corresponding to the NCSX plasma vessel. In a real device, of course,
some of these magnets will need to be removed in order to make space for 
access ports for diagnostic, heating, and fueling systems.

To evaluate the impact of ports on magnet arrangements, two more
sets of magnet arrangements were optimized, one with hexahedral geometry and
the other with curved brick geometry, this time excluding magnets
that collide with any of the ports foreseen for NCSX. Most of these ports
lie on the outboard side of the plasma vessel. As in the previous parameter
scans, each arrangement had a different value of the radial extent. 
The curved bricks were initialized with normal polarizations.

\begin{figure}
    \begin{center}
    \includegraphics[width=0.5\textwidth]{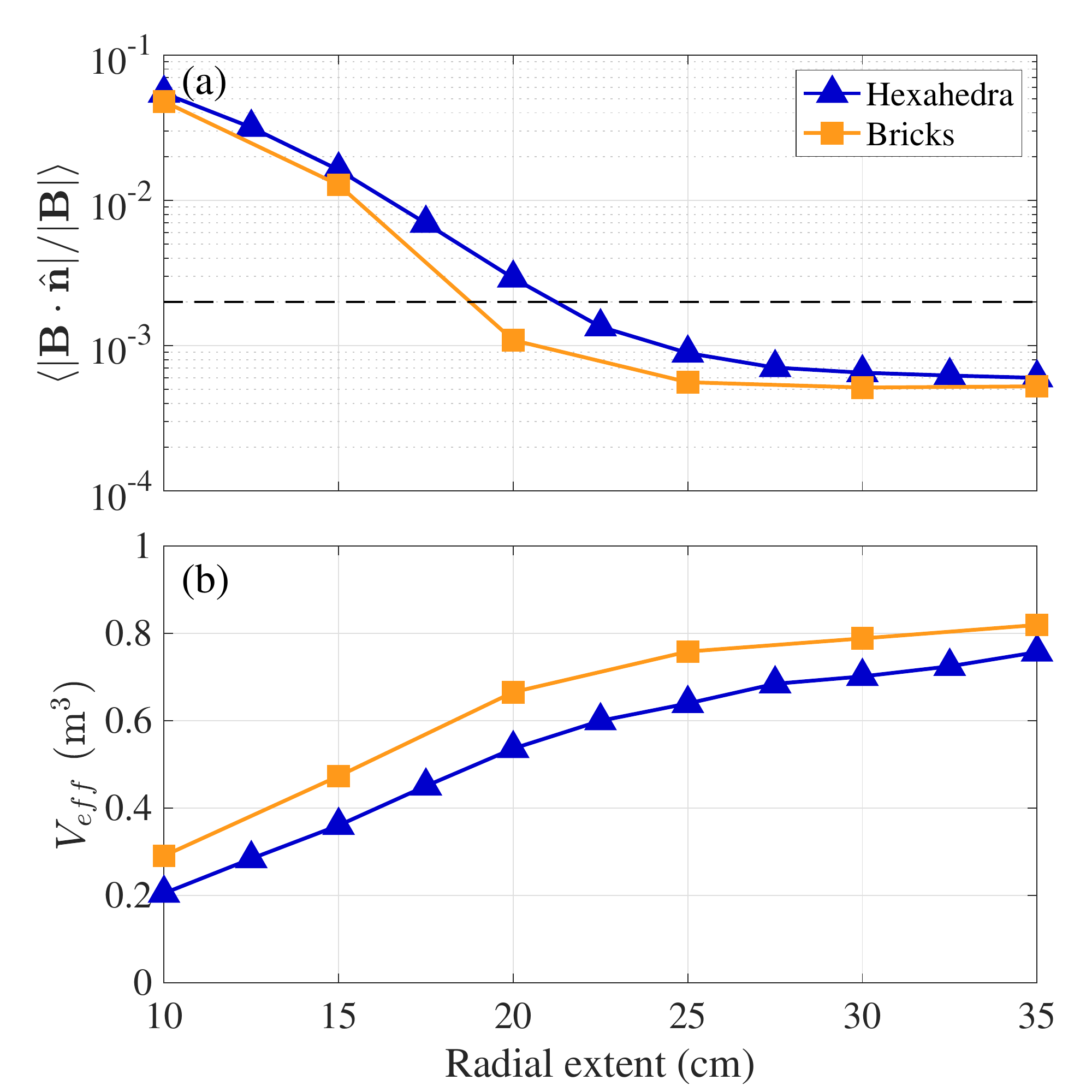}
    \caption{Values of (a) $\relbnorm$ and (b) $V_\text{eff}$ per half-module
             for magnet arrangements with hexahedral and curved brick geometry
             that left space for the NCSX ports.}
    \label{fig:bnorm_veff_ports}
    \end{center}
\end{figure}

\begin{figure*}
    \begin{center}
    \includegraphics[width=0.75\textwidth]{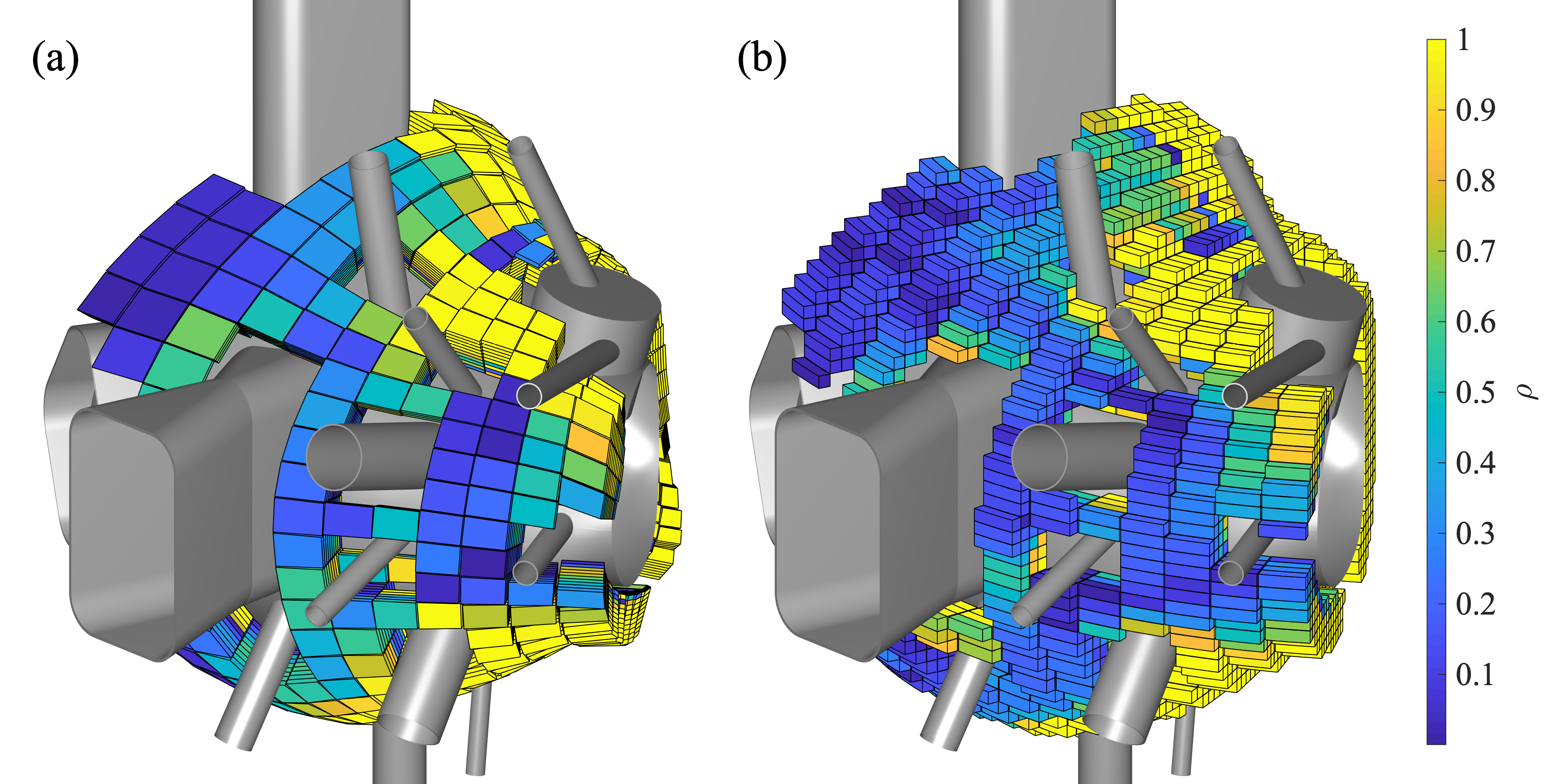}
    \caption{Renderings of magnet arrangements that leave space for the
             NCSX ports, viewed from the outboard side. 
             (a) Hexahedra with a radial extent of 22.5~cm;
             (b) Curved bricks with a radial extent of 20~cm.}
    \label{fig:magnets_ports}
    \end{center}
\end{figure*}

The attainable values $\relbnorm$ and $V_\text{eff}$ for the two parameter 
scans are shown in Fig.~\ref{fig:bnorm_veff_ports}. The lowest radial extent 
for which hexahedral
arrangements attained $\relbnorm<0.002$ was 22.5~cm, 5~cm higher than the
case without ports considered. Remarkably, the curved brick arrangements
attained $\relbnorm<0.002$ at the same radial extent as in the case without
ports, although $\relbnorm$ was about a factor of 2 higher. The 22.5~cm
hexahedral arrangement had $V_\text{eff}=0.60~\text{m}^3$ per half-period, 
whereas the 20~cm curved brick configuration had 
$V_\text{eff}=0.67~\text{m}^3$ per half-period; hence, the discrepancy in 
$V_\text{eff}$ between the hexahedral and brick arrangements was lower with
all NCSX ports considered. Renderings of each of these arrangements are
shown in Fig.~\ref{fig:magnets_ports}.

\subsection{Gaps for containment structures}
\label{ssect:gaps}

All magnet configurations considered thus far have exhibited very small gap
spacing between adjacent magnets ($\geq$0.5~cm between the faces of adjacent
hexahedra and $\geq$1~mm between the faces of adjacent curved bricks).
However, the magnet mounting concepts presently under consideration require
magnets to be enclosed in stainless-steel casings that are
in turn mounted to external structures (Fig.~\ref{fig:sas_ribs}). 
If such designs are ultimately 
adopted, larger gap spacings between magnets may be necessary in order
to leave room for the walls of the casings.

\begin{figure}
    \begin{center}
    \includegraphics[width=0.5\textwidth]{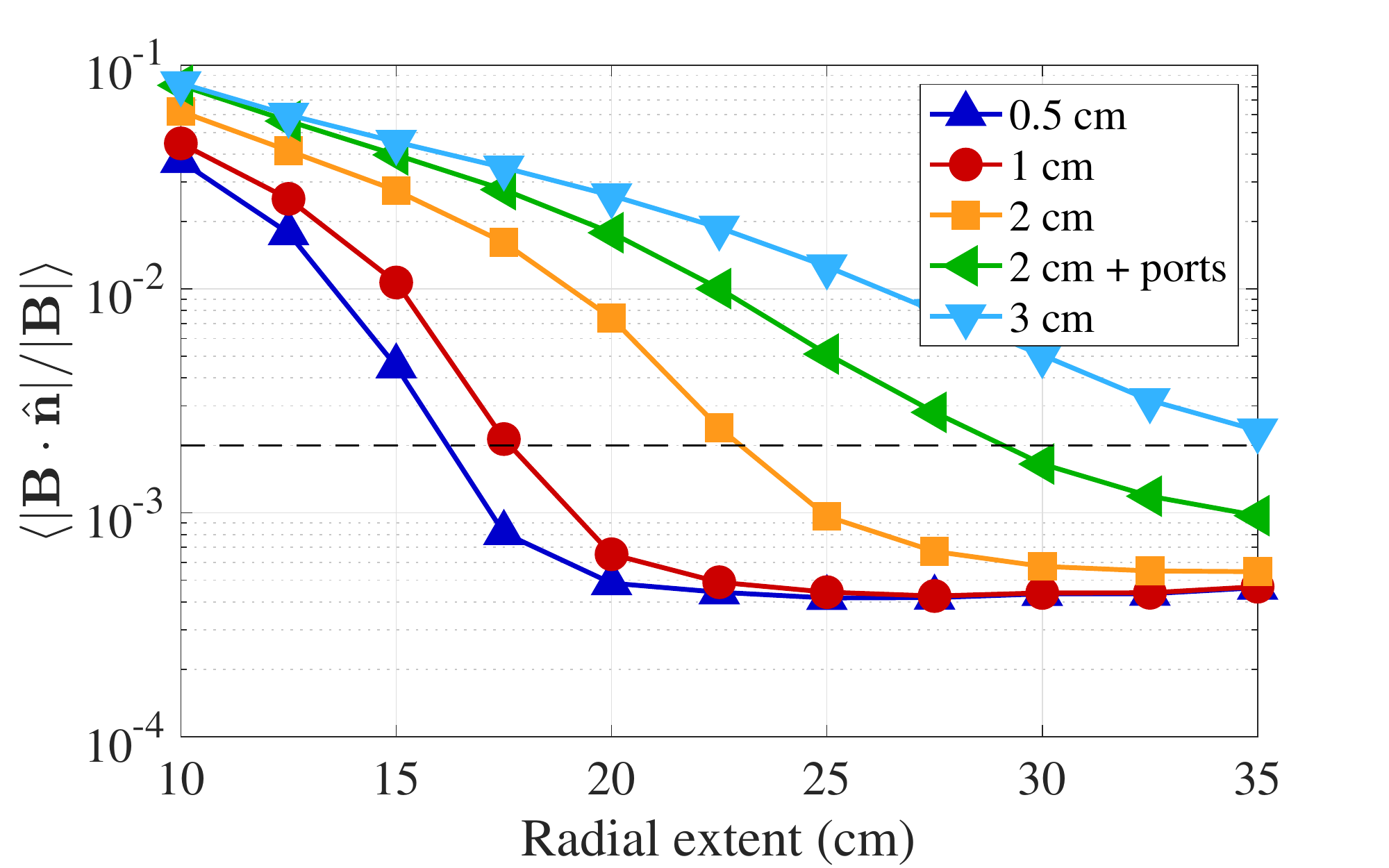}
    \caption{Values of $\relbnorm$ as a function of radial extent for 
             arrangements of hexahedral magnets with different minimum 
             gap spacings enforced between the lateral faces of adjacent
             hexahedra. For the case of 2~cm gaps, an additional arrangement
             was considered in which magnets colliding with the NCSX port set
             were removed.}
    \label{fig:bnorm_r_gaps}
    \end{center}
\end{figure}

Gap spacing requirements impact the amount of magnet material that can fill
a layer of a given thickness. Hence, an increase in the minimum gap
spacing would lead to a corresponding increase in the radial extent required
for the magnets to produce an adequate confining field. This effect is 
quantified for the case of hexahedral magnets in Fig.~\ref{fig:bnorm_r_gaps},
which shows the attainable $\relbnorm$ versus radial extent for arrangements
with increasing minimum gap spacings. The smallest value, 0.5~cm, was employed 
for the hexahedral arrangements studied in Sec.~\ref{sect:perp_axis} and 
\ref{sect:free_axis}. 

The first increment of the gap spacing, from 0.5~cm to 1~cm, 
requires a relatively small increase of approximately 2~cm in radial 
extent to satisfy the criterion for $\relbnorm$. Subsequent increments in the
gap spacing require much larger increases in the radial extent, however:
an increase from 1~cm to 2~cm in gap spacing calls for an increase of 
approximately 5~cm in radial extent, and an increase from 2~cm to 3~cm in
gap spacing calls for an increase of more than 10~cm in radial extent. One 
factor driving the nonlinearity of this relationship is that 3~cm is comparable 
to the magnet size in some portions of the arrangement.

Also shown in Fig.~\ref{fig:bnorm_r_gaps} is a version of the arrangement
with 2~cm gap spacing excluding magnets that overlap with the NCSX ports.
As with the case explored in Sec.~\ref{ssect:ports}, the removal of these
magnets results in an increase of the required radial extent, this time
to 30~cm.

The above study indicates that the hexahedral arrangement under consideration
is workable for gap spacings of up to at least 2~cm, which would leave ample 
room for each hexahedral magnet (or radial stack of magnets) to be enclosed in 
a casing with a standard wall thickness of 0.25~in (6.35~mm). To accommodate 
even larger gap spacing, it may be necessary to modify the parameters of the 
array; for example, by reducing the toroidal and poloidal grid cell resolution 
in order to boost the filling factor of magnet material between the limiting 
surfaces. 

Although the analysis in this section focused on the case of hexahedral
magnets, similar considerations would apply for arrangements of 
curved-brick magnets.
We also note that one need not enforce large spaces between every brick,
as multiple bricks may be grouped together within the same casing.

\section{Conclusions and future work}
\label{sect:conclusions}

In support of efforts to develop stellarators that employ arrays of permanent
magnets for plasma confinement, we have developed the \textsc{Magpie} code
to enable studies of different geometric concepts for magnets and facilitate
rapid fine-tuning of the arrangements according to physics and engineering
requirements. To date, two geometric concepts have been developed and
studied: curved bricks, which align to an arbitrary cylindrical grid, and 
quadrilaterally-faced hexahedra, which conform closely to toroidal plasma 
geometry.

The hexahedral concept was designed as a discrete implementation of a
magnetized layer with polarization constrained to be normal to a surface
surrounding the plasma. However, it was found that the physical limitations
on the strength of present-day rare-Earth magnets make it impossible for
normally-polarized hexahedra to attain adequate field accuracy, at least
for the targeted NCSX-like plasma configuration.

If the constraint for normal polarization is relaxed, however, both hexahedra
and curved bricks are capable of confining the target plasma configuration
with rare-Earth magnets. With both 
concepts, magnet layers conforming to the NCSX plasma vessel with thicknesses
on the order of 20~cm were shown to be capable of generating adequate magnetic
fields to confine the targeted plasma equilibria. For the hexahedral concept, 
an increase of 5~cm
in radial extent is required if magnets are removed to make space for all of
the NCSX ports. If larger gap spacings are required for magnet casings or 
other support structures, the minimum thickness will increase further.
In the case of hexahedral magnets, usable magnet arrays were found for
gap spacings of at least 2~cm.

The dipole moment distributions obtained by the optimizer had a noticeable
dependence on both the geometric properties of the magnet arrangement and
the initial conditions for the optimization procedure. Solutions for 
curved brick arrangements whose moments were initialized normally to the
toroidal limiting surface generally had lower effective magnet volumes than 
those for which the moments were initialized along the radial unit vector.
Furthermore, solutions for hexahedral arrangements tended to have lower 
effective volumes than solutions for curved brick arrangements.
Thus, while brick-shaped magnets have the advantage of geometric simplicity,
surface-conforming hexahedral magnets may admit configurations with lower
overall magnet quantities.

The immediate application for this work
is a stellarator experiment planned to be constructed at PPPL
using components from the NCSX experiment, including the vacuum vessel and 
the toroidal field coils. However, \textsc{Magpie} has been developed to 
aid in the design of permanent magnet arrays for any stellarator configuration.
This study has demonstrated the code's ability to facilitate conceptual
studies of different magnet geometries and to evaluate the trade-offs of
different design choices.

The findings presented here represent just the first steps in developing a
buildable permanent magnet array for a stellarator. Many critical questions
remain to be addressed in determining the overall feasibility of applying
permanent magnets to stellarators. This includes developing tolerance
criteria for magnet fabrication---both for geometry and magnetization---and 
assessing the associated costs to maintain adequate field accuracy.
In addition, modeling to date has not accounted for corrections to the
magnetization of each magnet due to the presence of backgroud fields; 
prototype studies will be needed to determine the corrections.
The mounting structures, of which some early concepts were shown in
this paper (Fig.~\ref{fig:sas_ribs}), must be qualified to withstand the
forces on each magnet---preliminary estimates indicate that a typical
hexahedral magnet will experience 5-10~kN due to the background magnetic field.
Furthermore, the construction of a large array consisting of multiple metric 
tons of magnetic material will require the development of suitable tooling and
assembly practices.
Efforts are currently underway to construct a prototype magnet array
in order to address these and other issues.

In support of these efforts, further refinements and additions to 
the \textsc{Magpie} code and the optimization approach are 
anticipated. One area of flexibility that we have
not yet explored in depth would be to reduce the effective volume by allowing
for nonuniformity in the thickness of the magnet layer, with thicker 
portions in the regions where large magnetic fields are required (particularly 
the inboard side) and thinner or nonexistent portions in less-critical areas.
Additional geometric concepts may be implemented in \textsc{Magpie} depending
on further research and development in magnet fabrication. 
In terms of the optimization, alternative techniques currently under
consideration would constrain each magnet to have one of a few discrete
polarization directions, rather than leaving the direction completely free
as is the case in the continuous quasi-Newton method employed in this work. 
Such constraints could greatly simplify the fabrication requirements.

Finally, the target plasma configuration studied in this paper
required a relatively thick magnet layer for good field accuracy despite its 
low average field strength of 0.5~T. This may call into question the utility
of permanent magnets if the target configuration were to be scaled up to
reactor-relevant parameters; in particular, with a much higher magnetic field.
However, as discussed in Ref.~\cite{zhu2020a}, the target configuration 
studied here was designed with modular coils in mind, and we expect that
it is possible to find alternative plasma configurations that would require
less magnetic material for a given level of field strength. 
To this end,
the \textsc{Magpie} code was designed to be relatively simple to include
in optimization loops by codes such as \textsc{Stellopt} \cite{stellopt},
which could be used to improve both the plasma equilibrium and the magnet
array geometry in order to reduce the volume of magnets required for adequate
confinement. Furthermore, as proposed in Ref.~\cite{helander2020a}, it would
also be worthwhile to investigate reactor concepts that use tilted planar 
coils or mildly non-planar coils in conjunction with permanent magnets, which
may still be simpler to construct than conventional stellarator reactor
designs that rely solely upon non-planar coils.

\section{Acknowledgments}

The authors would like to thank M.~Zarnstorff and R.~Mercurio for the helpful 
discussions.
This work was supported by the US Department of Energy under contract 
number DE-AC02-09CH11466. 
The digital data for this pper may be found at
\href{https://dataspace.princeton.edu/jspui/handle/88435/dsp01pz50gz45g}
{\textcolor{blue}
{https://dataspace.princeton.edu/jspui/handle/88435/dsp01pz50gz45g}}.

\appendix
\section{Details on the design of hexahedral arrays}
\label{app:qhex_construction}

\subsection{Base grid generation}
\label{ssect:base_grid}

The procedure for designing a hexahedral array in \textsc{Magpie} begins
with the definition of the volume in which magnets are permitted, specified
as the region between an outer toroidal limiting surface and an inner toroidal
limiting surface. The outer surface must enclose the inner surface, and both
are assumed to enclose the target plasma. To date, the inner surface has
been supplied as a set of Fourier harmonics for the $r$ and $z$ coordinates
as functions of poloidal angle $\theta$ and toroidal angle $\phi$, and
the outer surface has been implied through the definition of a uniform
radial extent.

With the magnet volume thus defined, the magnet design proceeds by generating 
a base grid.
The base grid consists of a two-dimensional array of vertices arranged on
the outer limiting surface. Each set of four adjacent vertices defines
a cell in which a magnet is to be placed. The base faces of each magnet are 
constrained not to extend behind any of its respective base grid vertices. 

The vertices (and, by extension, the bases of the magnets) 
may be arranged to interface
in a convenient way with their respective mounting structures. For the 
configuration shown in Fig.~\ref{fig:schematic_rendering}b, each poloidal
row of magnets was intended to be mounted onto a planar support structure
as shown in Fig.~\ref{fig:sas_ribs}. In order for the bases of the
magnets to line up along these planar structures, the vertex points in
each poloidal row of the base grid were chosen to have the same azimuthal
(toroidal) angle $\phi$.

\subsection{Definition of the bounding planes}
\label{ssect:bounding_planes}

Once the base grid is defined, the next step is to determine the bounding
planes between adjacent magnets. Since the magnets are ultimately oriented
such that their axes are perpendicular to the inner limiting surface,
the bounding planes should also intersect this surface at a 
near-perpendicular angle. To ensure this, the base grid vertices are first
projected onto the inner limiting surface 
such that each projection line intersects the surface normally.
This is accomplished by solving the system of three nonlinear equations,

\begin{equation}
    \mathbf{r}_v = \mathbf{r}_p(\theta,\phi) 
        + \ell\hat{\mathbf{n}}(\theta,\phi),
\end{equation} 

\noindent for $\ell$, $\theta$, and $\phi$. Here, $\theta$ and $\phi$ are 
the poloidal and toroidal angles, respectively, of the projected point 
$\mathbf{r}_p(\theta,\phi)$ on the inner limiting surface from which a 
perpendicular projection line intersects the vertex point $\mathbf{r}_v$.
The axis of this line is the unit normal vector to the inner limiting surface
$\mathbf{\hat{n}}(\theta,\phi)$, and $\ell$ is the distance along the axis
from $\mathbf{r}_p(\theta,\phi)$ to $\mathbf{r}_v$. The solution to these 
equations is not necessarily unique, especially for vertex points near 
concave regions of the limiting surface, so it is important to supply 
reasonable initial guesses of $\ell$, $\theta$, and $\phi$ to ensure 
continuity in the projection lines associated with adjacent vertex points.

\begin{figure}
    \begin{center}
    \includegraphics[width=0.3\textwidth]{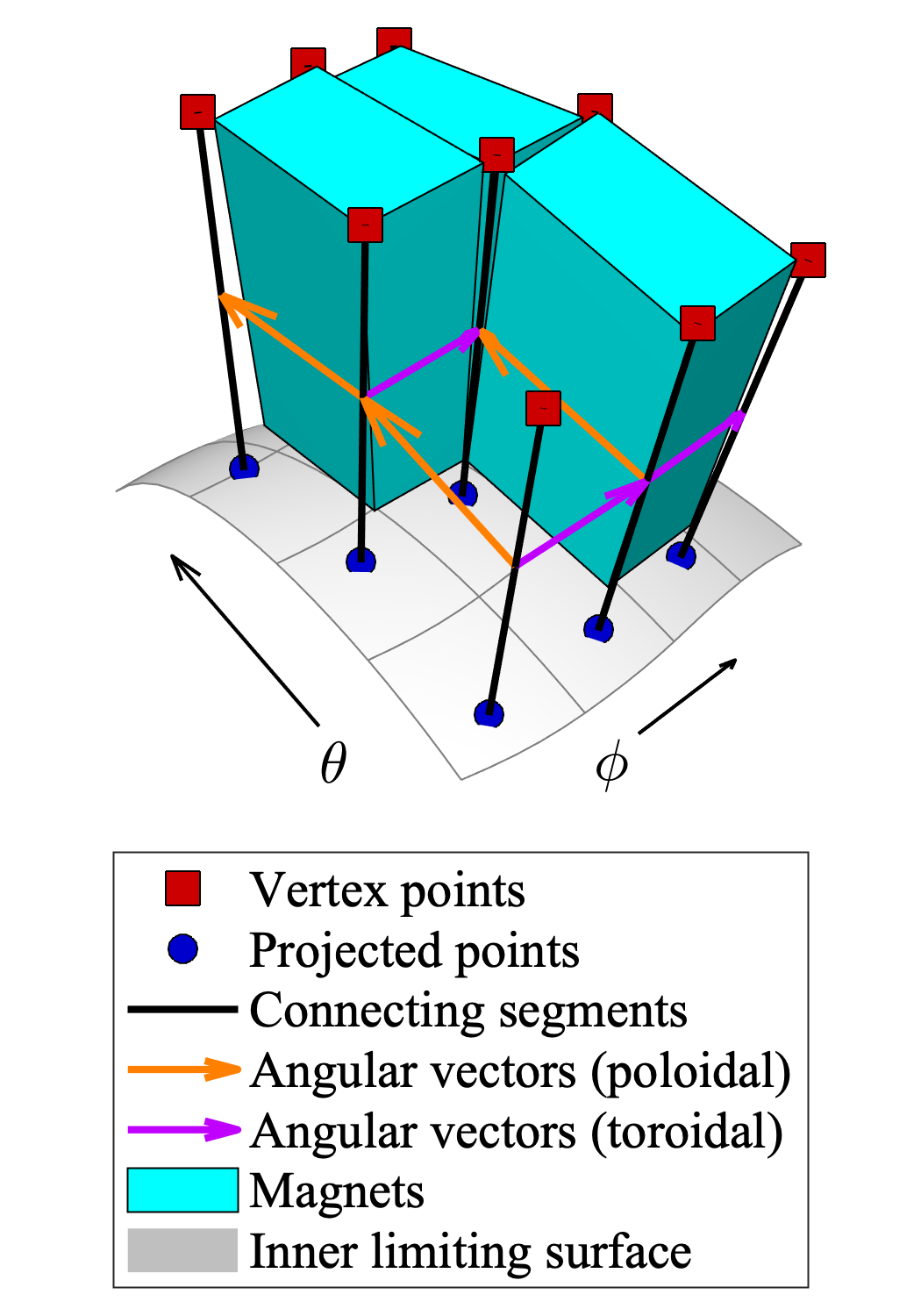}
    \caption{Points of reference for four cells of the base grid, including
             vertex points, projected points on the inner limiting surface, and
             their respective connecting segments. Radial vectors are parallel 
             to the connecting segments. Hexahedral magnets constructed from
             the grid are shown in three of the cells. Curves of constant
             poloidal angle $\theta$ and toroidal angle $\phi$ are also shown on
             the inner limiting surface.}
    \label{fig:bounding_plane_refs} 
    \end{center}
\end{figure}

The key reference points and directions for determining the bounding planes
are illustrated in Fig.~\ref{fig:bounding_plane_refs}.
Each pair of vertex points, projected points, and connecting segments demarks
a boundary between adjacent grid cells and is thus used to determine the 
corresponding bounding plane. The normal vector of each bounding plane is
derived from the cross-product of an angular vector (a vector between the 
midpoints of the connecting segments), and the radial vector (derived 
as an average of the axes of the connecting segments). The plane is fixed
to a reference point whose coordinates are calculated by averaging the
coordinates of the vertex and projected points. Some boxcar smoothing may be
applied to neighboring angular vectors, radial vectors, and plane reference
points in order to avoid sharp transitions between adjacent bounding planes.

With the bounding planes between adjacent grid cells defined, the lateral 
faces of the hexahedral magnets within the grid cells are then determined. The 
normal vector of each lateral face is that of the nearest bounding plane,
and the plane of the face is offset by a specified distance into the 
grid cell to enforce a finite gap spacing between the (parallel) face of the
adjacent magnet. The planes of each of the four lateral faces intersect
to form four radial edge lines. 

The magnet's axis is determined from 
the normal vector of the inner limiting surface at the average poloidal and
toroidal angles of the four projected points on the surface at the corners
of its grid cell.
The axis also defines the normal vectors of the top and base faces. The
top face is positioned to maintain a minimum separation distance from the
limiting surface. The base face is positioned not to not have a greater
elevation than any of the grid cell's vertex points along the magnet axis.

Following this procedure, each hexahedron will not collide with its four 
nearest neighbors due to the constraint that adjacent faces be parallel. 
However, it is still possible for overlaps to occur with other nearby hexahedra,
especially those across corners in the grid. If such overlaps are detected,
the lateral faces of the overlapping hexahedra are iteratively withdrawn
inwards until the overlap is resolved.

Depending on the input settings such as grid resolution, radial thickness,
and limiting surface geometry, it is possible for the above procedure to
result in over-constrained hexahedra with undesirable or erroneous properties. 
These could include, for example, opposite faces that intersect one another.
Magnets with such undesirable properties are identified and eliminated from
the output set. Since this will leave gaps in the magnet grid, it is 
advisable to adjust the input settings to minimize the number of erroneous
hexahedra. 

\subsection{Concave regions}
\label{ssect:concave_regions}

\begin{figure}
    \begin{center}
    \includegraphics[width=0.5\textwidth]{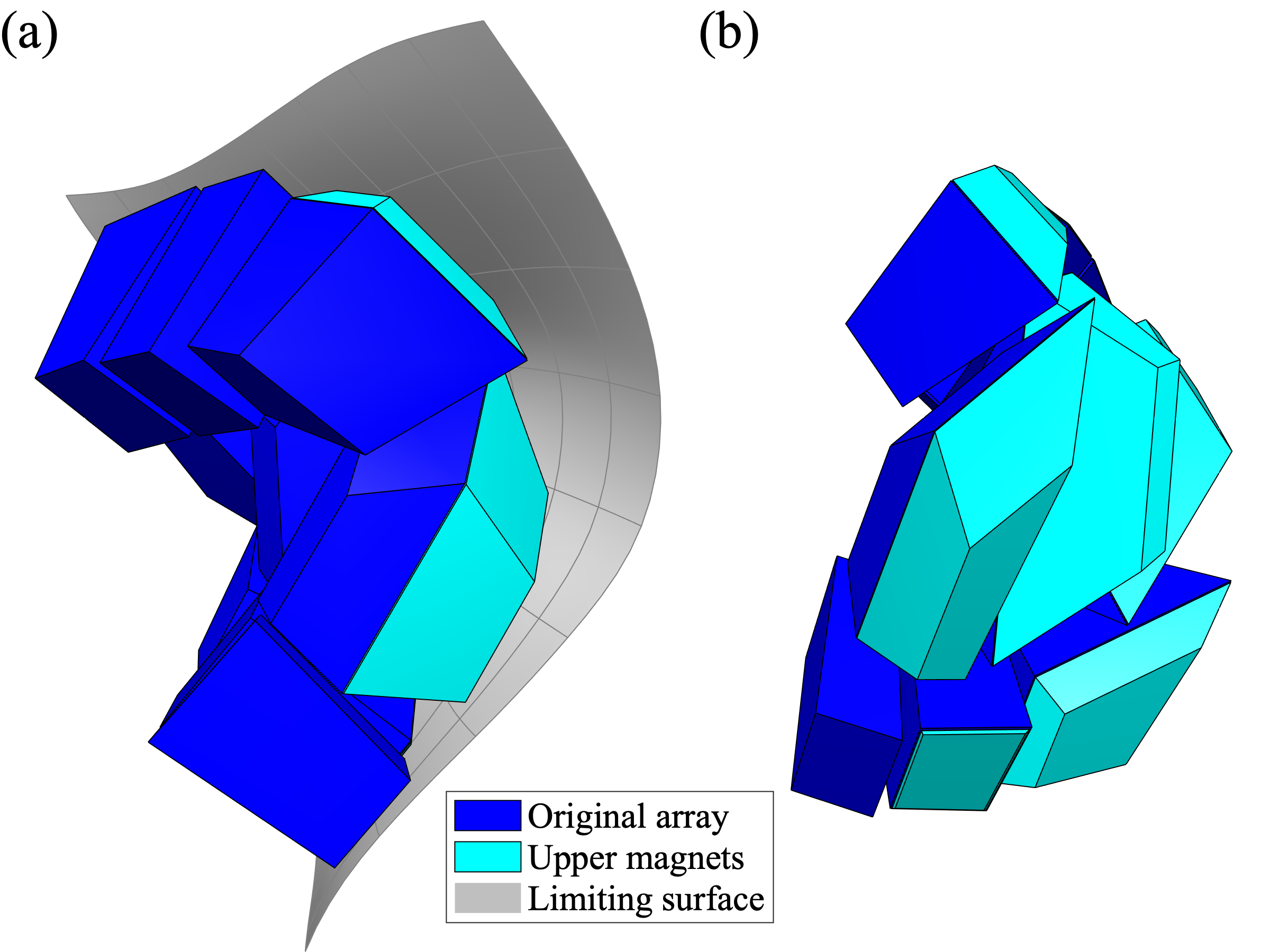}
    \caption{Subset of hexahedral magnets in the vicinity of a concave region
             of the limiting surface. Magnets in the original array are 
             shown in blue; upper magnets added to a subset of the grid cells
             are shown in cyan. (a) view from the inboard side of the torus;
             (b) view from inside the limiting surface.}
    \label{fig:upper_qhex}
    \end{center}
\end{figure}

In concave regions of the limiting surface, it is possible that significant
gaps could arise between the surface and the top face of the hexahedral
magnet. To fill this volume, the code offers the option to add an 
``upper'' hexahedral magnet between the original magnet and the vessel.
Examples of such upper magnets are shown in Fig.~\ref{fig:upper_qhex}.
The upper hexahedron's base face coincides with the original hexahedron's 
top face, and the upper hexahedron tapers along the axis as it expands into
the concave region of the limiting surface.

\bibliographystyle{unsrt}

\begin{thebibliography}{10}

\bibitem{mynick2006a}
H.~E. Mynick.
\newblock Transport optimization in stellarators.
\newblock {\em Physics of Plasmas}, 13:058102, 2006.

\bibitem{beidler1990a}
C.~Beidler, G.~Grieger, F.~Herrnegger, E.~Harmeyer, J.~Kisslinger, W.~Lotz,
  H.~Maassberg, P.~Merkel, J.~N{\"u}hrenberg, F.~Rau, J.~Sapper, F.~Sardei,
  F.~Scardovelli, A.~Schl{\"u}ter, and H.~Wobig.
\newblock {Physics and engineering design for Wendelstein VII-X}.
\newblock {\em Fusion Technology}, 17:148, 1990.

\bibitem{sapper1990a}
J.~Sapper and H.~Renner.
\newblock {Stellarator Wendelstein VII-AS: physics and engineering design}.
\newblock {\em Fusion Technology}, 17:62, 1990.

\bibitem{anderson1995a}
F.~S.~B. Anderson, A.~F. Almagri, D.~T. Anderson, P.~G. Matthews, J.~N.
  Talmadge, and J.~L. Shohet.
\newblock {The Helically Symmetric Experiment (HSX): goals, design and status}.
\newblock {\em Fusion Technology}, 27:273, 1995.

\bibitem{merkel1987a}
P.~Merkel.
\newblock Solution of stellarator boundary value problems with external
  currents.
\newblock {\em Nuclear Fusion}, 27:867, 1987.

\bibitem{landreman2017a}
M.~Landreman.
\newblock An improved current potential method for fast computation of
  stellarator coil shapes.
\newblock {\em Nuclear Fusion}, 57:046003, 2017.

\bibitem{helander2020a}
P.~Helander, M.~Drevlak, M.~Zarnstorff, and S.~C. Cowley.
\newblock Stellarators with permanent magnets.
\newblock {\em Physical Review Letters}, 124:095001, 2020.

\bibitem{aleksandrov2014a}
A.~Aleksandrov and A.~Menshov.
\newblock {Magnet design for the SNS Laser Stripping Experiment}.
\newblock In {\em {Proceedings of the 5th International Particle Accelerator
  Conference, Dresden, Germany}}, page TUPRO117, 2014.

\bibitem{thonet2016a}
P.~A. Thonet.
\newblock {Use of permanent magnets in multiple projects at CERN}.
\newblock {\em IEEE Transactions on Applied Superconductivity}, 26:4101404,
  2016.

\bibitem{hoffstaetter2017a}
G.~Hoffstaetter, D.~Trbojevic, and C.~Mayes.
\newblock {CBETA Design Report}.
\newblock Technical Report BNL-114549-2017-IR, Brookhaven National Laboratory,
  2017.

\bibitem{zhu2001a}
Z.~Q. Zhu and D.~Howe.
\newblock Halbach permanent magnet machines and applications: a review.
\newblock {\em IEEE Proceedings - Electric Power Applications}, 148:299, 2001.

\bibitem{roberson1989a}
C.~W. Roberson and P.~Sprangle.
\newblock A review of free-electron lasers.
\newblock {\em Physics of Fluids B}, 1:3, 1989.

\bibitem{oshea2010a}
F.~H. O'Shea, G.~Marcus, J.~B. Rosenzweig, M.~Scheer, J.~Bahrdt,
  R.~Weingartner, A.~Gaupp, and F.~Gr{\"u}ner.
\newblock Short period, high field cryogenic undulator for extreme performance
  x-ray free electron lasers.
\newblock {\em Physical Review Special Topics - Accelerators and Beams},
  13:070702, 2010.

\bibitem{bluemlich2008a}
B.~Bl{\"u}mlich, J.~Perlo, and F.~Casanova.
\newblock {Mobile single-sided NMR}.
\newblock {\em Progress in Nuclear Magnetic Resonance Spectroscopy}, 52:197,
  2008.

\bibitem{benabderrahmane2012a}
C.~Benabderrahmane, P.~Berteaud, M.~Vall{\'e}au, C.~Kitegi, K.~Tavakoli,
  N.~B{\'e}chu, A.~Mary, J.~M. Filhol, and M.~E. Couprie.
\newblock {Nd$_2$Fe$_{14}$B and Pr$_{2}$Fe$_{14}$B magnets characterization and
  modelling for cryogenic permanent magnet undulator applications}.
\newblock {\em Nuclear Instruments and Methods in Physics Research A}, 669:1,
  2012.

\bibitem{wang2012a}
J.-P. Wang, N.~Ji, X.~Liu, Y.~Xu, C.~S{\'a}nchez-Hanke, Y.~Wu, F.~M.~F.
  de~Groot, L.~Allard, and E.~Lara-Curzio.
\newblock {Fabrication of Fe$_{16}$N$_2$ films by sputtering process and
  experimental investigation of origin of giant saturation magnetization in
  Fe$_{16}$N$_2$}.
\newblock {\em IEEE Transactions on Magnetics}, 48:1710, 2012.

\bibitem{jiang2016a}
Y.~Jiang, M.~{Al Mehedi}, E.~Fu, Y.~Wange, L.~Allard, and J.-P. Wang.
\newblock {Synthesis of Fe$_{16}$N$_2$ compound free-standing foils with 20
  MGOe magnetic energy product by nitrogen ion-implantation}.
\newblock {\em Scientific Reports}, 6:25436, 2016.

\bibitem{nielson2010a}
G.~H. Nielson, C.~O. Gruber, J.~H. Harris, D.~J. Rej, R.~T. Simmons, and R.~L.
  Strykowsky.
\newblock {Lessons learned in risk management on NCSX}.
\newblock {\em IEEE Transactions on Plasma Science}, 38:320, 2010.

\bibitem{rummel2012a}
T.~Rummel, K.~Ri{\ss}e, G.~Ehrke, K.~Rummel, A.~John, T.~M{\"o}nnich, K.-P.
  Buscher, W.~H. Fietz, R.~Heller, O.~Neubauer, and A.~Panin.
\newblock {The superconducting magnet system of the stellarator Wendelstein
  7-X}.
\newblock {\em IEEE Transactions on Plasma Science}, 40:769, 2012.

\bibitem{bosch2013a}
H.-S. Bosch, R.~C. Wolf, T.~Andreeva, J.~Baldzuhn, D.~Birus, T.~Bluhm,
  T.~Br{\"a}uer, H.~Braune, V.~Bykov, A.~Cardella, F.~Durodi{\'e}, M.~Endler,
  V.~Erckmann, G.~Gantenbein, D.~Hartmann, D.~Hathiramani, P.~Heimann,
  B.~Heinemann, C.~Hennig, M.~Hirsch, D.~Holtum, J.~Jagielski, J.~Jelonnek,
  W.~Kasparek, T.~Klinger, R.~K{\"o}nig, P.~Kornejew, H.~Kroiss, J.~G. Krom,
  G.~K{\"u}hner, H.~Laqua, H.~P. Laqua, C.~Lechte, M.~Lewerentz, J.~Maier,
  P.~McNeely, A.~Messiaen, G.~Michel, J.~Ongena, A.~Peacock, T.~S. Pedersen,
  R.~Riedl, H.~Riemann, P.~Rong, N.~Rust, J.~Schacht, F.~Schauer, R.~Schroeder,
  B.~Schweer, A.~Spring, A.~St{\"a}bler, M.~Thumm, Y.~Turkin, L.~Wegener,
  A.~Werner, D.~Zhang, M.~Zilker, and {et al.}
\newblock {Technical challenges in the construction of the steady-state
  stellarator Wendelstein 7-X}.
\newblock {\em Nuclear Fusion}, 53:126001, 2013.

\bibitem{alderman2002a}
J.~Alderman, P.~K. Job, R.~C. Martin, C.~M. Simmons, and G.~D. Owen.
\newblock {Measurement of radiation-induced demagnetization of Nd-Fe-B
  permanent magnets}.
\newblock {\em Nuclear Instruments and Methods in Physics Research A}, 481:9,
  2002.

\bibitem{landreman2020a}
{M. Landreman et al.}
\newblock Calculation of permanent magnet arrangements for stellarators.
\newblock In preparation.

\bibitem{zhu2020a}
C.~Zhu, M.~Zarnstorff, D.~Gates, and A.~Brooks.
\newblock Designing stellarators using perpendicular permanent magnets.
\newblock {\em Nuclear Fusion}, 60:076016, 2020.

\bibitem{zhu2020b}
C.~Zhu, K.~C. Hammond, M.~Zarnstorff, T.~Brown, D.~Gates, K.~Corrigan,
  M.~Sibilia, and E.~Feibush.
\newblock Topology optimization of permanent magnets for stellarators.
\newblock {\em Nuclear Fusion}, in press [preprint:
  https://arxiv.org/abs/2005.05504].

\bibitem{zhu2017a}
C.~Zhu, S.~R. Hudson, Y.~Song, and Y.~Wan.
\newblock New method to design stellarator coils without the winding surface.
\newblock {\em Nuclear Fusion}, 58:016008, 2018.

\bibitem{zarnstorff2001a}
M.~C. Zarnstorff, L.~A. Berry, A.~Brooks, E.~Fredrickson, G.-Y. Fu,
  S.~Hirshman, S.~Hudson, L.-P. Ku, E.~Lazarus, D.~Mikkelsen, D.~Monticello,
  G.~H. Neilson, N.~Pomphrey, A.~Reiman, D.~Spong, D.~Strickler, A.~Boozer,
  W.~A. Coopern, R.~Goldston, R.~Hatcher, M.~Isaev, C.~Kessel, J.~Lewandowski,
  J.~F. Lyon, P.~Merkel, H.~Mynick, B.~E. Nelson, C.~N{\"u}hrenberg, M.~Redi,
  W.~Reiersen, P.~Rutherford, R.~Sanchez, J.~Schmidt, and R.~B. White.
\newblock {Physics of the compact advanced stellarator NCSX}.
\newblock {\em Plasma Physics and Controlled Fusion}, 43:A237, 2001.

\bibitem{nelson2003a}
B.~E. Nelson, L.~A. Berry, A.~B. Brooks, M.~J. Cole, J.~C. Chrzanowski, H.-M.
  Fan, P.~J. Fogarty, P.~L. Goranson, P.~J. Heitzenroeder, S.~P. Hirshman,
  G.~H. Jones, J.~F. Lyon, G.~H. Nielson, W.~T. Reiersen, D.~J. Strickler, and
  D.~E. Williamson.
\newblock {Design of the national compact stellarator experiment (NCSX)}.
\newblock {\em Fusion Engineering and Design}, 66-68:169, 2003.

\bibitem{halbach1980a}
K.~Halbach.
\newblock {Design of permanent multipole magnets with oriented rare Earth
  cobalt material}.
\newblock {\em Nuclear Instruments and Methods}, 169:1, 1980.

\bibitem{dahlgren2005a}
F.~Dahlgren, P.~Goranson, and P.~Titus.
\newblock {Structural analysis of the NCSX vacuum vessel}.
\newblock {\em Fusion Science and Technology}, 47:926, 2005.

\bibitem{pomphrey2001a}
N.~Pomphrey, L.~Berry, A.~Boozer, A.~Brooks, R.~E. Hatcher, S.~P. Hirshman,
  L.-P. Ku, W.~H. Miner, H.~E. Mynick, W.~Reiersen, D.~J. Strickler, and P.~M.
  Valanju.
\newblock Innovations in compact stellarator coil design.
\newblock {\em Nuclear Fusion}, 41:339, 2001.

\bibitem{hirshman1983a}
S.~P. Hirshman and J.~C. Whitson.
\newblock Steepest-descent moment method for three-dimensional
  magnetohydrodynamic equilibria.
\newblock {\em Physics of Fluids}, 26:3553, 1983.

\bibitem{hirshman1986a}
S.~P. Hirshman, W.~I. {van Rij}, and P.~Merkel.
\newblock {Three-dimensional free boundary calculations using a spectral
  Green's function method}.
\newblock {\em Computer Physics Communications}, 43:143, 1986.

\bibitem{nemov1999a}
V.~V. Nemov, S.~V. Kasilov, W.~Kernbichler, and M.~F. Heyn.
\newblock Evaluation of 1/{$\nu$} neoclassical transport in stellarators.
\newblock {\em Physics of Plasmas}, 6:4622, 1999.

\bibitem{stellopt}
S.~A. Lazerson, J.~Schmitt, C.~Zhu, J.~Breslau, and {STELLOPT Developers}.
\newblock {STELLOPT}.
\newblock https://doi.org/10.11578/dc.20180627.6, 2020.

\end{thebibliography}

\end{document}